\documentclass[article,nojss]{jss}
\usepackage{amsmath}
\usepackage{natbib}
\usepackage{amssymb}
\usepackage{booktabs}
\usepackage{subcaption}

\author{Jason B. Cho\\ Cornell University \And
        Daniel R. Kowal\\ Cornell University \And
        Haoxuan Wu\\ Headlands Technologies LLC \AND
        David S. Matteson\\ Cornell University \And
        Toryn L.J. Schafer\\ Texas A\&M University}
\title{dsp: An \proglang{R} Package for Bayesian Time Series Modeling with the Dynamic Shrinkage Process}

\Plainauthor{Jason B. Cho, Dan Kowal, Haoxuan Wu, David S. Matteson, Toryn L.J. Schafer}
\Plaintitle{An R Package for Bayesian Time Series Modeling with the Dynamic Shrinkage Process}
\Shorttitle{\pkg{dsp}: Bayesian Time Series Modeling with the Dynamic Shrinkage Process}

\Abstract{
\pkg{dsp} provides a unified state-space framework for modeling nonstationary time series using the Dynamic Shrinkage Process (DSP), a class of global-local shrinkage priors. DSP adaptively regularizes time-varying parameters to shrink noise while preserving abrupt changes and other local structure. The package supports Gaussian and count-data smoothing, changepoint detection, time-varying parameter regression, and volatility estimation, with posterior predictive draws available for missing-value imputation and forecasting. Nonstationary time series often contain periods of relative stability punctuated by abrupt changes in mean or variance. \pkg{dsp} offers a common interface for model specification, posterior computation, summarization, and visualization across models that were previously developed separately in the literature. Core functionality is implemented in \proglang{C++} via \pkg{Rcpp} for computational efficiency. We describe the underlying models, package design, and the main functions for fitting, summarizing, and plotting posterior results.
}

\Keywords{Bayesian Inference, Time Series Analysis, Gibbs Sampling, Markov Chain Monte Carlo (MCMC), State-Space Modeling}
\Plainkeywords{Bayesian Inference, Time Series Analysis, Gibbs Sampling, Markov Chain Monte Carlo (MCMC), State-Space Modeling}

\Address{
  Toryn L.J. Schafer \\
  Department of Statistics \\
  Assistant Professor \\
  Texas A\&M University \\
  3143 TAMU \\
  College Station, TX, 77843\\
  E-mail: \email{tschafer@tamu.edu}\\
  URL: \url{https://toryn.netlify.app/}
}

\begin{document}

\section{Introduction}

Global-local shrinkage priors have gained significant traction in Bayesian statistics for their ability to achieve adaptive regularization in high-dimensional models. By combining a single global parameter controlling overall shrinkage with local scale parameters, these priors reduce noise while preserving sharp local structures in the signal. This adaptability makes global-local shrinkage priors particularly useful for parameter regularization \citep{bhadra2019lasso}.

In particular, we focus on the dynamic shrinkage process (DSP) by \citet{dsp}, a flexible class of global-local priors designed for time series modeling. DSP generalizes existing global-local priors such as the horseshoe prior \citep{horseshoe_theory} and has already been extended to a broad set of problems, including trend filtering for Gaussian data and time-varying parameter regression \citep{dsp}, changepoint detection \citep{abco,dspDrift}, seasonal-trend decomposition \citep{bastion}, negative binomial Bayesian trend filtering for integer-valued time series \citep{dspCount}, and adaptive stochastic volatility (ASV) for time-varying volatility estimation \citep{asv}.

\pkg{dsp} is an \proglang{R} package that provides a unified, flexible, and computationally efficient framework for adaptive modeling. Performance-critical components are implemented in \proglang{C++} via \pkg{Rcpp}, enabling fast posterior computation and scalability to large data sets \citep{eddelbuettel2011}. The package offers a consistent interface for specifying models, running MCMC-based inference, and summarizing and visualizing results. Each fitted model returns full posterior predictive draws at every observation time, so missing-value imputation and forecasting are available directly from the MCMC output. We demonstrate its use through synthetic and empirical examples, highlighting its practical advantages for trend estimation, changepoint detection, and volatility modeling.

\subsection{Related Work}

\pkg{dsp} consolidates the DSP-based Bayesian time series models into a single, coherent programmatic framework \citep{dsp, dspCount, dspDrift, abco}. In addition to the dynamic horseshoe prior (DHS), \pkg{dsp} also implements several related prior families, including the horseshoe \citep[HS;][]{horseshoe_theory}, Bayesian LASSO \citep[BL;][]{Bayesianlasso}, stochastic volatility \citep[SV;][]{SVdsp}, and normal-inverse-gamma \citep[NIG;][]{nig}. Among existing state-space modeling packages such as \pkg{bsts} \citep{bsts}, \pkg{dlm} \citep{dlm}, and \pkg{KFAS} \citep{kfas}, \pkg{dsp} is one of the few state-space modeling packages that directly embeds global-local shrinkage. A comparable \proglang{R} package is \pkg{shrinkTVP} \citep{shrinkageTVP}, which relies on the triple-gamma prior \citep{tgamma}, but is restricted to time-varying parameter regression. In contrast, \pkg{dsp} provides a unified interface for a broader range of time series models and priors.

Related functionality outside \proglang{R} is limited. In \proglang{MATLAB}, the package accompanying \citet{korobolisbayesian2022} implements a range of shrinkage and sparsity priors for regression, vector autoregressive, factor and time-varying parameter models, in the latter case by applying static or independent shrinkage priors to the stacked coefficient vector rather than by modeling the prior scales as a stochastic process. In \proglang{Python}, state-space models are provided by \pkg{statsmodels} \citep{statsmodels} while \pkg{PyMC} \citep{pymc} supports user-specified Bayesian models. To our knowledge, however, no maintained library in either environment implements the dynamic shrinkage process, so that \pkg{dsp} is currently its only general-purpose implementation.

Additionally, we chose to implement model-specific samplers rather than utilize a general-purpose Bayesian sampling engine such as \proglang{Stan} \citep{stan}, \proglang{JAGS} \citep{jags}, \proglang{BUGS} \citep{bugs}, or \pkg{NIMBLE} \citep{nimble}. Our choice is primarily driven by the following two reasons. First, closed-form full conditional distributions are available for the majority of the parameters and latent variables \citep{dsp, dspCount, abco}. Furthermore, the conditional independence structure of the state equation renders the precision matrix of the latent state banded, which our sampler exploits to draw the entire state exactly from its joint full conditional with compute time scaling linearly in the sample size \citep{simSmooth}. Second, the geometry of the posterior of our model specification is generally unsuitable as written for Hamiltonian Monte Carlo (HMC) samplers such as \proglang{Stan}. Specifically, global-local shrinkage priors place a heavy-tailed prior on a scale parameter that multiplies the quantity being shrunk, which induces a funnel-shaped posterior in which the feasible range of the location contracts as the scale approaches zero. This geometry is a property of the prior rather than of any particular sampling scheme, and it is known to be difficult for gradient-based samplers, whose step size must accommodate the most constrained direction \citep{hmcHier, reparam}. Accordingly, \citet{regHorseshoe} provide alternative parameterizations of the horseshoe prior for settings in which a direct implementation encounters sampling difficulties. The DSP compounds this difficulty, since the local log-scales follow an autoregressive process rather than being independent, so that the funnel recurs at each time point and the scales are serially dependent. An additional consideration arises for the changepoint model. Specifically, the autoregressive coefficient of the log-volatility process switches according to a binary indicator determined by an unknown threshold, so that the log posterior is piecewise constant in that threshold and gradient-based inference would require either marginalization or methods adapted to discontinuous targets \citep{dhmc}. Taken together, the availability of closed-form full conditionals and the difficulty of the posterior geometry make model-specific samplers the more practical choice for this family of models.

\section[Methods]{Methods}\label{sec:methods}
\subsection{Dynamic Shrinkage Process}
Let $\{y_t\}_{t=1}^{T}$ denote an observed univariate time series of length $T$, and let $\{\omega_t\}_{t=1}^{T}$ be a time-varying parameter of interest associated with $\{y_t\}_{t=1}^{T}$. The DSP is a flexible family of global-local shrinkage priors defined by the hierarchical form:
\begin{equation}
    \begin{aligned}
    &[\omega_{t}|v_{t}] \stackrel{indep}{\sim} N(0,\exp(v_{t})), \\
    &v_{t} = m + \psi_t + \eta_{t}, \\
    &[\eta_{t}|\alpha,\beta] \stackrel{iid}{\sim}Z(\alpha,\beta,0,1),
    \end{aligned}\label{eq:dsp_g}
\end{equation}
where $m$ is the global scale parameter and $\psi_{t} + \eta_{t}$ is the local scale parameter. Similar to other global-local shrinkage priors, $\eta_{t}$ is an independent and identically distributed local parameter across $t = 1, \ldots, T$ from a heavy-tailed distribution, allowing selective deviation from strong shrinkage. Specifically, $\eta_{t}$ follows the $Z$-distribution parameterized as $Z(\alpha,\beta,m_{z},\sigma_{z})$ with the probability density function:
\begin{align*}
&f(z|\alpha,\beta,m_{z},\sigma_z) = \{\sigma_{z}B(\alpha,\beta)\}^{-1} \exp\{(z-m_{z})/\sigma_{z}\}^{\alpha}[1+\exp\{(z-m_{z})/\sigma_{z}\}]^{-(\alpha+\beta)} & z \in \mathbb{R},
\end{align*}
where $B(\alpha,\beta)$ denotes the Beta function.

Many existing shrinkage priors emerge as special cases of the $Z$-distribution \citep{barndorff1982normal}. Specifically, several well-known shrinkage priors defined on the variance scale correspond exactly to a $Z$-distribution on the log-scale for particular choices of $\alpha$ and $\beta$; examples include the Strawderman--Berger ($\alpha = \frac{1}{2}, \beta = 1$; \citep{strawderman}), normal-exponential-gamma ($\alpha = 1, \beta = c-2$ for shape parameter $c>2$; \citep{griffin2005}), and normal-Jeffreys ($\alpha = \beta \to 0$; \citep{figueiredo}) priors. Notably, choosing $\alpha = \beta = \frac{1}{2}$ leads to $\exp(\eta_{t})$ following an inverted Beta distribution. By additionally setting $\psi_t = 0$ and $m \sim Z(1/2,1/2,0,1)$, DSP simplifies exactly to the Horseshoe prior \citep{horseshoe_theory}.

The distinctive feature of DSP is its incorporation of time dependence between the local terms, $\psi_t$, enabling dynamic modeling of the shrinkage parameter, hence the name \textit{Dynamic Shrinkage Process}. In the original formulation by \citet{dsp}, the term $\psi_t$ follows a first-order autoregressive structure. The additional choice of $\alpha = \beta = \frac{1}{2}$ yields the dynamic horseshoe prior (DHS):
\begin{equation}
    \begin{aligned}
        &v_{t} = m + \phi(v_{t-1} - m)  + \eta_t, &[\eta_{t}] \stackrel{iid}{\sim}Z(\frac{1}{2},\frac{1}{2},0,1),
    \end{aligned}\label{eq:dhs_ar1}
\end{equation}
Following \citet{horseshoe_theory}, \pkg{dsp} assigns $m\sim Z(\frac{1}{2},\frac{1}{2},0,1)$. \pkg{dsp} also assumes the shifted Beta prior $(\phi+1)/2 \sim \mathrm{Beta}(10,2)$ \citep{dsp}, which enforces stationarity and anticipates moderate to strong positive autocorrelation in $v_t$. Beyond DHS, \pkg{dsp} also implements four alternative prior families: the Horseshoe \citep[HS;][]{horseshoe_theory}, Bayesian Lasso \citep[BL;][]{Bayesianlasso}, SV \citep[SV;][]{SVdsp}, and Normal-Inverse-Gamma \citep[NIG;][]{nig}.

For time-varying parameter regression (Section~\ref{sec:g_regression}) and changepoint detection (Section~\ref{sec:g_cp}), the temporal dependence structure of the local scale process differs from the AR(1) specification in Equation~\ref{eq:dhs_ar1}. Specifically, \pkg{dsp} uses a diagonal VAR(1) specification for time-varying parameter regression and a lag-1 threshold SV model, TSV(1), for changepoint detection. These model-specific dependence structures are separate from the choice of prior family, and their detailed parameterizations are given in the corresponding sections.

\subsection{Bayesian Trend Filter for Gaussian Data}\label{sec:g_smoothing}
As one of many applications of DSP, \cite{dsp} propose a Bayesian Trend Filter (BTF) for adaptive smoothing. The motivation derives from the following (random walk plus noise) dynamic linear model, which is widely used to model and filter noisy time series data \citep{west}:
\begin{equation*}
    \begin{aligned}
    & y_{t} = \mu_{t} + \epsilon_{t}, & &[\epsilon_t|\sigma_{\epsilon}] \stackrel{iid}{\sim} N(0,\sigma^2_{\epsilon}), \\
    &\omega_{t} = \mu_{t} - \mu_{t-1}, & & [\omega_{t}|\sigma_{\omega}] \stackrel{iid}{\sim} N(0,\sigma^2_{\omega}).
\end{aligned}
\end{equation*}
The behavior of the increments, $\omega_t$, governs the evolution of $\mu_{t}$: when $\omega_{t}$ is near zero, $\mu_{t}$ is locally constant, while large $\omega_{t}$ corresponds to large changes in $\mu_{t}$. The random walk model includes a single parameter, $\sigma_\omega$, to control the evolution. To enhance the model's flexibility, various extensions have been proposed that impose constraints on $\omega_{t}$. Trend filtering extends this idea by replacing the first-order difference with a higher-order difference, $\omega_t = \Delta^D \mu_{t}$, where $D$ denotes the order of differencing. The choice of $D$ determines the local structure of the estimated trend. $D=1$ encourages piecewise constant behavior, $D=2$ encourages piecewise linear behavior, and higher values of $D$ allow higher-order piecewise polynomial behavior. To obtain adaptive smoothing, penalties or priors are then placed on $\omega_t$ so that most differenced increments are shrunk toward zero while a small number are allowed to remain large. \citet{l1tf1} and \citet{l1tf2} introduce the $\ell_1$ trend filter, which penalizes the $\omega_t$ using an $\ell_1$ norm to enforce sparsity in $\omega_t$, inducing piecewise polynomial structure in $\mu_t$. Alternatively, \citet{SVfilter} takes a fully Bayesian approach by placing priors such as the HS, BL, or NIG priors on $\omega_t$; these are also implemented in \pkg{dsp}.

We consider the extension introduced by \citet{dsp}, which assigns the DHS prior to the differenced mean parameter $\omega_t$. The resulting model is the Bayesian Trend Filter with the DHS prior (BTF-DHS):
\begin{equation}
\begin{aligned}
      &y_{t} = \mu_{t} + \epsilon_{t}, \quad & & [\epsilon_t|\sigma_{\epsilon,t}] \stackrel{indep}{\sim} N(0,\sigma^2_{\epsilon,t}), \\
    &\omega_{t} = \Delta^{D}\mu
    _t, \quad & &[\omega_{t}] \sim \mathrm{DHS},
\end{aligned}
\label{eq:BTF}
\end{equation}
where $\mathrm{DHS}$ represents the Dynamic Horseshoe prior, specified by Equation~\ref{eq:dsp_g} with the AR(1) process in Equation~\ref{eq:dhs_ar1} for the log-variance of $\omega_t$. Note that $\sigma^2_{\epsilon,t}$ in Equation~\ref{eq:BTF} is time-varying, generalizing the constant variance case to accommodate heteroskedasticity. \pkg{dsp} supports three specifications for $\sigma^2_{\epsilon,t}$: constant variance, stochastic volatility \citep[SV;][]{SVop,kimetal}, and adaptive stochastic volatility \citep[ASV;][]{asv}, the latter two detailed in Section~\ref{sec:asv}.

Compared with the Gaussian random walk model, BTF-DHS provides two important advantages. First, it introduces temporal dependence in the shrinkage process through the additional time-varying term $\psi_t$, allowing the model to dynamically adjust the degree of regularization over time. Second, by leveraging a sparse, heavy-tailed prior for $\eta_t$, DSP provides improved adaptability in capturing both smooth and abrupt trend changes.

\subsection{Changepoint Detection Model}\label{sec:g_cp}
DSP provides a general framework for introducing temporal dependence in shrinkage parameters. In the DHS prior, this dependence is modeled using an AR(1) process as shown in Equation~\eqref{eq:dhs_ar1}. For changepoint detection, \citet{abco} replace this AR(1) dependence with a threshold autoregressive SV process, leading to the Adaptive Bayesian Changepoints with Outliers (ABCO) model. ABCO is designed to identify temporal locations associated with abrupt changes in the local mean while accounting for isolated outliers and time-varying volatility. This model has been used to quantify changes in COVID-19 transmission associated with social distancing interventions~\citep{covid_cp}.

ABCO decomposes the observed series into three components: a local mean $\{\mu_t\}$, a sparse additive outlier signal $\{\zeta_t\}$, and a heteroskedastic noise term $\{\epsilon_t\}$:
\begin{equation}
    \begin{aligned}
        &y_{t} = \mu_{t} + \zeta_{t} + \epsilon_t, \quad & [\epsilon_t|\sigma_{\epsilon,t}] \stackrel{indep}{\sim} N(0,\sigma_{\epsilon,t}^2). \\
    \end{aligned}\label{eq:abco}
\end{equation}
A threshold variant of DSP is imposed on the $D$th (usually for D = 1, or 2) differencing of the conditional mean parameter, $\{\mu_{t}\}$,  which is the primary variable of interest.
\begin{equation}
    \begin{aligned}
        &\omega_t  = \Delta^{D} \mu_{t},&\quad & [\omega_t|v_t] \stackrel{indep}{\sim} N(0,\exp(v_{t})), \\
        &v_{t} = m + (\phi_1 + \phi_2s_{t-1})(v_{t-1} - m) + \eta_t, & \quad &[\eta_{t}]\stackrel{iid}{\sim}Z(\frac{1}{2},\frac{1}{2},0,1).
    \end{aligned}\label{eq:abco_p}
\end{equation}
In the previous specification \eqref{eq:dhs_ar1}, the time dependence shrinkage parameter $\psi_t$ is modeled using first-order autoregression. In order to incorporate changepoints in the model, a threshold stochastic volatility model, TSV(1), is imposed on $\{v_t\}$ with $Z$-distributed innovations. This structure allows asymmetric response, $\phi_1$ versus $\phi_1 + \phi_2$, in $v_{t}$. Specifically, for some threshold $\gamma$,
\[
    s_{t}=\begin{cases}
			1, & \text{if $\log(\omega^2_{t}) > \gamma$}\\
            0, & \text{otherwise}
		 \end{cases}.
\]
For modeling the noise process $\{\sigma_{\epsilon,t}^2\}$, ABCO adopts a first-order SV model as proposed by \citet{SVop,kimetal} to accommodate potential heteroskedasticity; \pkg{dsp} also supports ASV \citep{asv} as an alternative specification for modeling the variances. In addition, ABCO's observation equation in~\eqref{eq:abco} incorporates an additive outlier process $\{\zeta_t\}$ to capture substantial deviations at specific time points. We assume the additive outliers are sparse and independent in time. Therefore, we employ an independent horseshoe+ shrinkage prior as proposed by \citet{horseshoe_P}. Following \citet{abco}, the outlier process is defined for $t = D+1, \ldots, T$, and $\zeta_t = 0$ at the first $D$ time points, which carry the initial states of $\{\mu_t\}$. Given the observed data, $\{\zeta_t\}$ strongly concentrates around zero, except at points where deviations from the mean are significantly larger in magnitude than the surrounding trend estimate. This also helps separate the additive outlier process from the flexible trend model.

\subsection{B-Spline Smoothing}\label{sec:g_spline}
DSP can also be used to estimate the conditional mean for irregularly spaced time series or functional data. To do so, we build on \citet{dspfunc}, who propose a DSP-based approach for estimating regression coefficient functions in scalar-on-function regression models. In this framework, the conditional mean function is represented as
\[
f(t) = \sum_{k=1}^{K} \beta_k \varphi_k(t),
\]
where \(\{\varphi_k\}_{k=1}^K\) is a collection of B-spline basis functions and \(\beta_1,\ldots,\beta_K\) are the associated basis coefficients. DSP is then placed on the \(D\)th-order differences of the coefficients to obtain a locally adaptive estimate of the conditional mean:
\begin{equation*}
\begin{aligned}
    &y_t = f(t) + \epsilon_t,  & &[\epsilon_t \mid \sigma_{\epsilon}] \stackrel{iid}{\sim} N(0,\sigma_\epsilon^2),\\
    &\omega_k = \Delta^D \beta_k, & &[\omega_k] \sim \mathrm{DHS}.
\end{aligned}
\end{equation*}
This basis expansion allows the conditional mean to be evaluated at unequally spaced inputs. In the \pkg{dsp} package, an order 4 B-spline basis is used, so each \(\varphi_k(t)\) is a cubic basis function. Replacing the DHS prior with a normal-inverse-gamma prior recovers the Bayesian P-spline formulation of \citet{psplines}.

\subsection{Bayesian Trend Filter for Count Data}\label{sec:nb_smoothing}
Non-stationary count time series, which often exhibit abrupt shifts and fluctuations in trend, arise in various scientific domains, including biophysics \citep{losey2022}, epidemiology \citep{covidcount, count_ep,nbmodel}, and emergency services \citep{emergency}. To flexibly model such data, \citet{dspCount} propose a state-space framework that places the DSP on the $D$th-order differences of a latent log-intensity. The model is
\begin{equation}
\begin{aligned}
      & [y_t \mid \mu_t, r] \sim \mathrm{NB}\!\left(r,\, \frac{e^{\mu_t + o_t}}{r + e^{\mu_t + o_t}}\right),
        & & E[y_t \mid \mu_t, r] = e^{\mu_t + o_t}, \\
      & \omega_t = \Delta^{D} \mu_t,
        & & [\omega_t] \sim \mathrm{DHS},
\end{aligned}
\label{eq:nbBTF}
\end{equation}
where, for $t = 1, \dots, T$, $\mu_t$ is the latent log-intensity, $o_t$ is a known offset, and $r > 0$ is a scalar overdispersion parameter.

The resulting model, referred to as the Negative Binomial Bayesian Trend Filter (NB-BTF), offers a flexible approach for analyzing non-stationary count time series, particularly those characterized by low counts and structural breaks. The offset $o_t$ accommodates known observation-specific exposures (e.g., $o_t = \log N_t$ for population size $N_t$), so that $e^{\mu_t}$ represents a rate per unit exposure rather than an absolute count; setting $o_t = 0$ for all $t$ (the default) recovers the model on raw counts. The inclusion of $r$ allows the model to account for overdispersion. As $r \to \infty$, the negative binomial distribution converges to the Poisson distribution, recovering the standard Poisson likelihood. Thus, NB-BTF can be interpreted as a more general and robust extension of Poisson-based models for count data.

\subsection{Time-varying Parameter Regression}\label{sec:g_regression}
Time-varying parameters or coefficients are commonly introduced in economic and financial applications, as empirical data in these fields often exhibit structural changes driven by endogenous and exogenous shocks, systemic innovations, shifts in market conditions, fiscal policy interventions, and other evolving factors \citep{tvpreview}. The locally adaptive shrinkage property of DSP makes it particularly well-suited for modeling such time-varying parameters. In this context, we consider a time-varying parameter regression model with $p$ dynamic predictors, $\boldsymbol{x}_t = (x_{1,t},\ldots,x_{p,t})' \in \mathbb{R}^p$:
\begin{equation}
\begin{aligned}
      &y_t = \boldsymbol{x}_{t}^{'}\boldsymbol{\beta}_{t} + \epsilon_t,
      \quad &&[\epsilon_t|\sigma_{\epsilon,t}] \stackrel{indep}{\sim} N(0,\sigma_{\epsilon,t}^2), \\
      &\omega_{j,t} = \Delta^D \beta_{j,t},
      \quad &&[\omega_{j,t}] \sim \mathrm{DHS},
\end{aligned}
\label{eq:TVR}
\end{equation}
for $j=1,\ldots,p$, where $\boldsymbol{\beta}_t = (\beta_{1,t},\ldots,\beta_{p,t})' \in \mathbb{R}^p$ and $\omega_{j,t}$ denotes the $D$th-order difference of the $j$th time-varying coefficient with a DHS prior specified independently across $j$ through the coefficient-specific log-variance process $v_{j,t}$, with global scale parameter $m_j$, autoregressive parameter $\phi_j$, and innovations $\eta_{j,t}$.

\subsection{Modeling Time-Varying Variance}\label{sec:asv}
Volatility is often a primary object of interest alongside the conditional mean, especially in financial applications \citep{SVop,ARCH,GARCH2}. For the Gaussian trend filtering (Section~\ref{sec:g_smoothing}), changepoint detection (Section~\ref{sec:g_cp}), and time-varying parameter regression models (Section~\ref{sec:g_regression}), \pkg{dsp} allows the observation error variance $\sigma_{\epsilon,t}^2$ to be specified as either constant or time-varying.

A direct application of this idea is the Adaptive Stochastic Volatility (ASV) model by \citet{asv}, which places DSP on the \(D\)th-order differences of the log-variance process \(h_t\). This extends the Random Walk Stochastic Volatility (RWSV) model~\citep{SVRW1,SVRW2} by allowing the volatility process to be smooth over stable periods while remaining locally adaptive to abrupt changes. For a univariate mean-zero time series \(y_t\), \(t=1,\ldots,T\), ASV is defined as
\begin{equation}
    \begin{aligned}
        &y_t = \exp(h_t/2)\varepsilon_t, & \quad &\varepsilon_t \stackrel{iid}{\sim} N(0,1),\\
        &\omega_t = \Delta^D h_t, & \quad &[\omega_t] \sim \mathrm{DHS}.
    \end{aligned}
\end{equation}
Through the global-local shrinkage prior, ASV effectively captures significant shifts and sudden changes within the volatility process. In addition to ASV, users may specify either a constant variance model with a normal-inverse-gamma prior or a SV model \citep{SVop, kimetal}, the latter implemented via the \pkg{stochvol} package \citep{stochvol}.

\section{Package Overview}
The \pkg{dsp} package is organized around model specification, posterior computation, and posterior analysis. A user first calls \code{dsp\_spec()} to specify the likelihood, model type, evolution-error prior, and any model-specific options. The resulting specification object is passed to \code{dsp\_fit()}, which performs MCMC sampling and returns a \code{dsp} object. The object can then be inspected with accessor functions for its components, \code{summary()} for a compact numerical summary, and \code{plot()} for visualization.

\subsection{Model Specification}
The \code{dsp\_spec()} function serves as the starting point to the \pkg{dsp} package. Users must specify two required parameters: \texttt{family} and \texttt{model}. The \texttt{family} argument accepts either \texttt{"gaussian"} for Gaussian or \texttt{"negbinomial"} for negative binomial models. For the Gaussian family, four model types are supported:
\begin{itemize}
    \item \texttt{"smoothing"}: Bayesian trend filter (Section~\ref{sec:g_smoothing}),
    \item \texttt{"changepoint"}: Changepoint detection with ABCO (Section~\ref{sec:g_cp}),
    \item \texttt{"bspline"}: B-Spline smoothing (Section~\ref{sec:g_spline}),
    \item \texttt{"regression"}: Time-varying parameter regression (Section~\ref{sec:g_regression}).
\end{itemize}
For the Negative Binomial family, the \texttt{"smoothing"} model described in Section~\ref{sec:nb_smoothing} is available. The required arguments of \code{dsp\_spec()} are \texttt{family} and \texttt{model}. For the \texttt{"regression"} model, the design matrix \texttt{X} must also be specified. All other arguments are optional and have default values. Their detailed descriptions are listed below.
\paragraph{Shared Arguments Across Models}
\begin{itemize}
    \item \texttt{D}: (integer scalar) Degree of differencing for $\omega_t$. Supported values are \texttt{0}, \texttt{1}, and \texttt{2}. Defaults to \texttt{2}.
    \item \texttt{obsSV}: (character) Specifies the model for the observation error variance, $\sigma^2_{\epsilon,t}$. Options include:
        \begin{itemize}
            \item \texttt{"const"}: Constant variance (default),
            \item \texttt{"SV"}: Stochastic volatility,
            \item \texttt{"ASV"}: Adaptive stochastic volatility.
        \end{itemize}
        For Gaussian \texttt{bspline} models, only \texttt{"const"} is supported. The options \texttt{"SV"} and \texttt{"ASV"} are available only for Gaussian \texttt{smoothing}, \texttt{changepoint}, and \texttt{regression} models. This argument is not available for negative binomial models.
    \item \texttt{evol\_error}: (character) Prior distribution for the evolution error $\omega_t$. Options include:
        \begin{itemize}
            \item \texttt{"DHS"}: Dynamic horseshoe (default),
            \item \texttt{"HS"}: Horseshoe,
            \item \texttt{"BL"}: Bayesian lasso,
            \item \texttt{"SV"}: Stochastic volatility,
            \item \texttt{"NIG"}: Normal-inverse-gamma.
        \end{itemize}
        This argument is not used by the ABCO changepoint model, which uses a model-specific prior (Equation~\ref{eq:abco_p}) for changepoint detection.
    \item \texttt{D\_asv}: (integer scalar) Degree of differencing for the log-variance process $h_t = \log(\sigma^2_{\epsilon,t})$. Supported values are \texttt{0}, \texttt{1}, and \texttt{2}. Defaults to \texttt{1}. Only used when \code{obsSV = "ASV"} for Gaussian \texttt{smoothing}, \texttt{changepoint}, and \texttt{regression} models.
    \item \texttt{evol\_error\_asv}: (character) Prior distribution for the evolution error in the log-variance process $h_t = \log(\sigma^2_{\epsilon,t})$. Options are the same as those for \texttt{evol\_error}. Defaults to \texttt{"HS"}. Only used when \code{obsSV = "ASV"} for Gaussian \texttt{smoothing}, \texttt{changepoint}, and \texttt{regression} models.
\end{itemize}
\paragraph{Model-Specific Arguments}
\begin{itemize}

  \item \texttt{bspline} (Gaussian):
  \begin{itemize}
    \item \texttt{times}: (length $T$ numeric vector, or \texttt{NULL}) Observation indices. If \texttt{NULL} (default), time points are assumed to be equally spaced.
    \item \texttt{num\_knots}: (integer scalar) Number of knots in the B-spline basis. Defaults to 20; must be at least 4.
  \end{itemize}

    \item \texttt{regression} (Gaussian):
    \begin{itemize}
        \item \texttt{X}: ($T \times p$ numeric matrix) Time-varying predictors required for fitting a regression model.
  \end{itemize}

  \item \texttt{changepoint} (Gaussian):
  \begin{itemize}
    \item \texttt{useAnom}: (logical) If \texttt{TRUE}, includes an additive anomaly component in the observation equation to capture abrupt outliers. Defaults to \texttt{FALSE}.
  \end{itemize}

  \item \texttt{smoothing} (Negative Binomial):
  \begin{itemize}
     \item \texttt{r\_user}: (numeric scalar); if ``NULL'' (Default), MCMC sampling is done for the overdispersion parameter \code{r}; If a positive numeric value is passed, the overdispersion is fixed to that value. Observation distribution approximately Poisson if a large numeric value is passed.
    \item \texttt{offset}: (length $T$ numeric vector) Offsets for the log-conditional expectation, used to adjust the model for exposure or baseline rate.
  \end{itemize}
\end{itemize}
\begin{table}[ht]
\centering
\begin{tabular}{lll}
\toprule
\textbf{Family} & \textbf{Model} & \textbf{Arguments} \\
\midrule
\texttt{gaussian} & \texttt{smoothing} & \texttt{D}, \texttt{evol\_error}, \texttt{obsSV},  \texttt{D\_asv}, \texttt{evol\_error\_asv} \\
\texttt{gaussian} & \texttt{bspline}  & \texttt{D}, \texttt{evol\_error}, \texttt{obsSV}, \texttt{times}, \texttt{num\_knots} \\
\texttt{gaussian} & \texttt{regression} & \texttt{D}, \texttt{evol\_error}, \texttt{obsSV},  \texttt{X},  \texttt{D\_asv}, \texttt{evol\_error\_asv}\\
\texttt{gaussian} & \texttt{changepoint} & \texttt{D}, \texttt{useAnom}, \texttt{obsSV},  \texttt{D\_asv},  \texttt{evol\_error\_asv} \\
\texttt{negbinomial} & \texttt{smoothing} & \texttt{D}, \texttt{evol\_error}, \texttt{r\_user}, \texttt{offset} \\
\bottomrule
\end{tabular}
\caption{Summary of shared and model-specific arguments by model type and family for the \code{dsp\_spec()} function.}
\label{tab:model-args}
\end{table}

Table~\ref{tab:model-args} summarizes the arguments of \code{dsp\_spec()} for each model type and family. The output of \code{dsp\_spec()} is a \code{dsp\_spec} object that records the selected observation family, model type, and associated model arguments. This object is passed to \code{dsp\_fit()} for posterior computation.

\subsection{MCMC Sampler}
The object instantiated via the \code{dsp\_spec()} function is then passed to the \code{dsp\_fit()} function for MCMC sampling. Additional arguments for \code{dsp\_fit()} are described below.
\begin{itemize}
    \item \texttt{y}: (length $T$ numeric vector) Observed time series. Missing values (\texttt{NA}) are allowed.
    \item \texttt{model\_spec}: (class \code{dsp\_spec}) Model specification generated from \code{dsp\_spec()}.
    \item \texttt{nsave}: (integer scalar) Number of MCMC iterations to record. Defaults to 1000.
    \item \texttt{nburn}: (integer scalar) Number of MCMC iterations to discard as burn-in. Defaults to 1000.
    \item \texttt{nskip}: (integer scalar) Number of MCMC iterations to skip between saved draws, i.e., save every $(\texttt{nskip} + 1)$th draw. Defaults to 4.
    \item \texttt{computeDIC}: (logical) If \texttt{TRUE} (default), compute the deviance information criterion (DIC) and the effective number of parameters $p_d$.
    \item \texttt{verbose}: (logical) If \texttt{TRUE}, print progress information to the console. Defaults to \texttt{FALSE}.
\end{itemize}
\subsection{Model Output}\label{sec:output_des}
The fitted \code{dsp} object is returned as a named list containing:
\begin{itemize}
    \item \texttt{mcmc\_output}: (named list) Posterior MCMC draws, with one named element per model parameter (e.g., \texttt{mu}, \texttt{ypred}, \texttt{obs\_sigma\_t2}). The set of parameters present depends on the chosen model and options; the descriptions below list all possibilities.
    \item \texttt{DIC}: (named list with elements \texttt{DIC} and \texttt{p\_d}, each of length 2) Deviance information criterion (DIC) and effective number of parameters. Lower values of \texttt{DIC} indicate a better tradeoff between fit and complexity for models fit to the same data. Two versions are reported, corresponding to the two standard definitions of the effective number of parameters, $p_D$ \citep{spiegelhalter2002} in the first entries and $p_V$ \citep{gelman2014} in the second entries. We recommend $p_D$ by default, since it is the more numerically stable of the two \citep{gelman2014}, and we use it throughout this article.
    \item \texttt{mcpar}: (named numeric vector) MCMC configuration parameters: \texttt{nsave} saved draws, \texttt{nburn} burn-in iterations, and \texttt{nskip} iterations discarded between saved draws, so that the thinning interval is $\texttt{nskip} + 1$.
    \item \texttt{model\_spec}: (class \code{dsp\_spec}) The model specification object used to fit the model.
\end{itemize}
These components are not intended to be accessed directly. Instead, \pkg{dsp} provides the following accessor functions:
\begin{itemize}
    \item \code{par_names(object)}: names of the parameters with stored posterior draws. These are the values accepted by the \code{pars} argument of \code{fitted()}, \code{summary()}, and \code{as.mcmc()}, and by the \code{type} argument of \code{plot()}.
    \item \code{fitted(object, pars = "mu", probs)}: pointwise posterior mean and quantiles of one parameter, returned as a matrix with one row per time point and quantile columns named as by \code{quantile()}. Regression coefficients \code{beta} are returned as an array with one slice per predictor, and a scalar parameter as a named vector. The default \code{probs = c(0.025, 0.25, 0.5, 0.75, 0.975)}. Simultaneous credible bands are returned by \code{credible_bands()}, described in Section~\ref{sec:bands}.
    \item \code{DIC(object)}: a data frame with one row for each of the $p_D$ and $p_V$ versions of the DIC, giving the DIC and the corresponding effective number of parameters.
    \item \code{get_spec(object)}: the \code{dsp\_spec} object used to fit the model.
    \item \code{nobs(object)}: the number of time points $T$, including any missing observations, which are imputed by the sampler.
\end{itemize}
Printing a fitted object reports its model specification and MCMC settings. Detailed descriptions of the parameters that may be returned by \code{par_names()} are given below. Each parameter is stored across \texttt{nsave} MCMC draws; the dimensions reported below refer to a single posterior draw.
\paragraph{Shared Parameters}
\begin{itemize}
    \item \texttt{mu}: (length $T$ numeric vector) Conditional mean of $y_t$.
    \item \texttt{ypred}: (length $T$ numeric vector) Posterior predictive $\tilde{y}_t$.
    \item \texttt{evol\_sigma\_t2}: (length $T$ numeric vector for \texttt{smoothing}/\texttt{changepoint}; length $p$ numeric vector for \texttt{bspline}; $T \times p$ numeric matrix for \texttt{regression}) Evolution error variance $\exp(v_t)$, i.e., the variance of $\omega_t$. For $D > 0$, the first $D$ entries are the variances of the initial states.
    \item \texttt{obs\_sigma\_t2}: (numeric scalar when the variance is constant in time, i.e., for \texttt{obsSV = "const"} and for \texttt{bspline}; otherwise length $T$ numeric vector) Observation variance $\sigma^2_{\epsilon,t}$ (\texttt{family = "gaussian"}).
    \item \texttt{dhs\_phi}: (numeric scalar; length $p$ numeric vector for \texttt{regression}) DHS AR(1) coefficient $\phi$ (\texttt{evol\_error = "DHS"} only).
    \item \texttt{dhs\_mean}: (numeric scalar; length $p$ numeric vector for \texttt{regression}) DHS unconditional mean $m$ (\texttt{evol\_error = "DHS"} only).
\end{itemize}

\paragraph{Model-Specific Parameters}
\begin{itemize}
    \item \texttt{changepoint} (Gaussian):
    \begin{itemize}
        \item \texttt{omega}: (length $T - D$ numeric vector) $D$th-order differences of $\mu_t$.
        \item \texttt{zeta}: (length $T$ numeric vector) Anomaly component (when \texttt{useAnom = TRUE}), equal to zero at the first $D$ time points.
        \item \texttt{zeta\_sigma\_t2}: (length $T$ numeric vector) Variance parameters of the anomaly component $\zeta_t$ (when \texttt{useAnom = TRUE}), equal to zero at the first $D$ time points.
        \item \texttt{gamma}: (numeric scalar) Threshold parameter in the thresholded DHS log-volatility dynamics.
    \end{itemize}
    \item \texttt{regression} (Gaussian):
    \begin{itemize}
        \item \texttt{beta}: ($T \times p$ numeric matrix) Time-varying regression coefficients.
    \end{itemize}
    \item \texttt{bspline} (Gaussian):
    \begin{itemize}
        \item \texttt{beta}: (length $p$ numeric vector) B-spline basis coefficients, where $p$ is the number of basis functions.
    \end{itemize}
    \item \texttt{smoothing} (Negative Binomial):
    \begin{itemize}
        \item \texttt{r}: (numeric scalar) Overdispersion parameter.
    \end{itemize}
\end{itemize}
\subsection{Summary}
A compact summary of a fitted model is obtained using the \code{summary()} function:
\begin{itemize}
    \item \texttt{object}: (class \code{dsp}) Fitted model object returned by \code{dsp\_fit()}.
    \item \texttt{pars}: (character vector) Names of parameters to summarize. Defaults to all parameters returned by \code{par_names(object)}.
    \item \texttt{probs}: (length 2 numeric vector) Probabilities of the posterior interval reported for scalar parameters. Defaults to \texttt{c(0.025, 0.975)}.
    \item \texttt{ess}: (logical) If \texttt{TRUE}, also report the minimum and median pointwise effective sample size of each time-varying parameter. Defaults to \texttt{FALSE}, since this requires one spectral estimate per time point and is slow for long series.
\end{itemize}
The result is an object of class \code{summary.dsp} with a dedicated \code{print()} method. It reports the model specification, the MCMC settings, the DIC, and one line per parameter. A scalar parameter is summarized by its posterior mean, standard deviation, interval, and effective sample size \citep{coda}. A time-varying parameter is summarized by its length and the range of its pointwise posterior means, so that the summary fits on one screen regardless of $T$; regression coefficients are reported with one line per predictor. The first $D$ evolution variances, which belong to the initial states and have their own prior, are reported on a separate line from the remaining evolution variances. Pointwise summaries of a time-varying parameter are instead returned by \code{fitted()}.
\subsection{Plot}\label{sec:plot}
The \code{plot()} function visualizes posterior inference results for objects of class \code{dsp}. The plotting behavior depends on the dimension of the posterior samples of the parameter named by \texttt{type}:
\begin{itemize}
    \item \textbf{1D (scalar parameters)}: a density-style summary is produced using a histogram with an overlaid kernel density estimate. The posterior mean and equal-tailed \((1-\alpha)100\%\) credible interval are marked. If \texttt{true\_values} is provided, the true value is overlaid.
    \item \textbf{2D (time-varying parameters)}: a time series plot is produced showing the posterior mean together with equal-tailed pointwise credible intervals. If \texttt{include\_joint\_bands = TRUE} and the parameter is one of \texttt{zeta}, \texttt{omega}, \texttt{ypred}, or \texttt{mu}, the simultaneous credible bands of Section~\ref{sec:bands} are also displayed. If \texttt{true\_values} is provided, the values are overlaid. Optionally, the observed data can also be added through \texttt{y\_obs}.
    \item \textbf{3D (multi-parameter time-varying quantities)}: a multi-panel collection of time series plots is produced, one panel for each slice of the third dimension. Each panel shows the posterior mean, pointwise credible intervals, optional joint bands when supported, and optional ground-truth values. The panel layout can be controlled through \texttt{nr} and \texttt{nc}; otherwise, a near-square layout is chosen automatically.
\end{itemize}
By default, pointwise credible intervals are constructed from the posterior samples using equal-tailed empirical quantiles at levels \(\alpha/2\) and \(1-\alpha/2\) at each time point.

Key arguments of the \code{plot()} function include:
\begin{itemize}
    \item \texttt{x}: (class \code{dsp}) Fitted model object returned by \code{dsp\_fit()}.
    \item \texttt{type}: (character) Name of the parameter to visualize; must be one of \code{par_names(x)}. Defaults to \texttt{"mu"}. For \texttt{evol\_sigma\_t2}, the first $D$ entries, which are the variances of the initial states, are removed before plotting. A constant observation variance is drawn as a horizontal band over time, so that it can be compared with a time-varying variance.
    \item \texttt{true\_values}: (numeric scalar, length $T$ numeric vector, or $T \times p$ numeric matrix) Optional ground-truth values to overlay. The required shape matches the plotted parameter.
    \item \texttt{y\_obs}: (length $T$ numeric vector) Optional observed data, available for 2D parameters.
    \item \texttt{times}: (length $T$ numeric vector, or \texttt{NULL}) Observation time points used as the x-axis. If \texttt{NULL}, $T$ equally spaced points on $[0,1]$ are used.
    \item \texttt{include\_joint\_bands}: (logical) If \texttt{TRUE}, overlay the simultaneous credible bands of Section~\ref{sec:bands}. Currently supported for the time-varying parameters \texttt{zeta}, \texttt{omega}, \texttt{ypred}, and \texttt{mu} (2D outputs), and \texttt{beta} (3D outputs).
    \item \texttt{alpha}: (numeric scalar) Credibility level used to construct posterior intervals and bands. Defaults to \texttt{0.05}.
    \item \texttt{xlab}, \texttt{ylab}, \texttt{main}: (character) Labels for the x-axis (\texttt{xlab}), y-axis (\texttt{ylab}), and plot title (\texttt{main}).
    \item \texttt{xlim}, \texttt{ylim}: (length 2 numeric vector) Optional x-axis and y-axis limits passed to the plotting routine.
    \item \texttt{mar}: (length 4 numeric vector) Optional margin specification passed to \texttt{par()}.
    \item \texttt{par\_args}: (list) Optional additional graphical parameters passed to \texttt{par()}, such as \texttt{cex.axis}, \texttt{cex.lab}, \texttt{cex.main}, \texttt{las}, or \texttt{mgp}.
    \item \texttt{legend}: (logical) If \texttt{TRUE}, add a legend to the plot. Defaults to \texttt{TRUE}.
    \item \texttt{legend\_cex}: (numeric scalar) Scaling factor for legend text size. Defaults to \texttt{1}.
    \item \texttt{legend\_pt\_cex}: (numeric scalar) Scaling factor for legend symbol size. Defaults to \texttt{2}.
    \item \texttt{nr}, \texttt{nc}: (integer scalars) Optional numbers of rows and columns for the multi-panel layout in 3D plots.
    \item \texttt{cp\_thres}: (numeric scalar) Cutoff proportion used to flag changepoints in fitted changepoint models. Annotations are added only when \texttt{model = "changepoint"}, the parameters \texttt{omega} and \texttt{gamma} are among \code{par_names(x)}, and the plotted parameter is one of \texttt{mu}, \texttt{ypred}, or \texttt{omega}.
\end{itemize}

\subsection{Simultaneous Credible Bands}\label{sec:bands}
The pointwise credible intervals returned by \code{fitted()} and drawn by \code{plot()} are computed from the posterior quantiles at each time point separately. When the trajectory of a parameter is of interest as a whole, \code{credible_bands()} instead returns simultaneous credible bands, which are calibrated jointly over all time points:
\begin{itemize}
  \item \texttt{object}: (class \code{dsp}) Fitted model object returned by \code{dsp\_fit()}.
  \item \texttt{pars}: (character) Name of one time-varying parameter; must be one of \code{par_names(object)}. Defaults to \texttt{"mu"}.
  \item \texttt{level}: (numeric scalar) Credibility level $1-\alpha$ of the bands. Defaults to \texttt{0.95}.
  \item \texttt{times}: (optional vector of length $T$) Labels for the time points of the series, such as dates. Defaults to \texttt{1:T}.
\end{itemize}
The bands are constructed as in \citet{crainiceanu2007}. At each time point, the band is centered at the posterior mean of the parameter, and its half-width is a multiple of the posterior standard deviation at that time point. The multiplier is the same at every time point and is chosen so that the band contains the entire trajectory in a proportion $1-\alpha$ of the posterior draws. The band is therefore wider than the pointwise intervals at the same credibility level. These are the bands displayed by \code{plot()} with \code{include_joint_bands = TRUE}, and \citet{simultaneous} compare them with alternative constructions.

The function also returns the simultaneous band scores (SimBaS) of \citet{simbas}. The SimBaS at time $t$ is the smallest $\alpha$ for which the $1-\alpha$ simultaneous band excludes zero at $t$. A small value is evidence that the parameter is nonzero at $t$, and because the band is calibrated over all time points, this evidence accounts for examining all time points at once. The smallest SimBaS over all time points is a global Bayesian $p$-value for the hypothesis that the parameter is zero at every time point. The scores are most useful for parameters for which zero is a value of interest. Two examples are the regression coefficients \texttt{beta} and the increments \texttt{omega}, for which a nonzero value marks a change in the level of the mean when $D = 1$ or in its slope when $D = 2$.

Printing the result reports the global Bayesian $p$-value and the periods in which the band excludes zero. The result itself is an object of class \code{dsp_bands}, a data frame with one row per time point and columns \texttt{time}, \texttt{mean}, \texttt{lower}, \texttt{upper}, and \texttt{simBaS}, and \code{as.data.frame()} returns it as a plain data frame. For \texttt{beta}, an additional column \texttt{predictor} identifies the regression coefficient. The \texttt{time} column is taken from \texttt{times}, and the $T-D$ increments \texttt{omega} are labelled with the last $T-D$ time points. Time points with no posterior variation, such as fixed initial values, are excluded when the band is calibrated. Examples are given in Sections~\ref{sec:abco_ex} and~\ref{sec:count_reg}.

\subsection{CODA Coercion}\label{sec:coda}
Posterior draws from a fitted \code{dsp} object can be coerced to an \code{mcmc} object, the class used by the \pkg{coda} package \citep{coda}, via an \code{as.mcmc()} method for class \code{dsp}, making the convergence-diagnostic and plotting functions in \pkg{coda} directly available. The generic is re-exported by \pkg{dsp}, so \code{as.mcmc()} is available without attaching \pkg{coda}.

\begin{itemize}
  \item \code{x}: (class \code{dsp}) Fitted model object returned by \code{dsp\_fit()}.
  \item \code{pars}: (character vector) Names of parameters to retain. Defaults to every parameter returned by \code{par_names(x)}.
\end{itemize}

Each retained parameter is flattened into an \code{nsave}-row matrix. Specifically, a scalar parameter is represented by a single column, while a time-varying or multi-dimensional parameter represents one column per index, named accordingly (e.g., \code{"mu[3]"}, \code{"beta[3, 2]"}). \code{DIC} and \code{p\_d} are posterior summaries rather than draws and are always excluded from the  \code{mcmc} object.

The returned object records the sampler's true iteration index: its start, end, and thinning interval are set from the \code{nburn}, \code{nsave}, and \code{nskip} used in the fit, and are returned by the \pkg{coda} functions \code{start()}, \code{end()}, and \code{thin()}.

The functions of \pkg{dsp} and \pkg{coda} are therefore complementary. The accessor functions and \code{summary()} work at the level of the model, reporting the specification, fitted values, and DIC, while draw-level diagnostics and intervals, such as \code{effectiveSize()}, \code{traceplot()}, \code{gelman.diag()}, and \code{HPDinterval()}, are applied to the coerced draws. An example of this coercion is shown in Subsection~\ref{sec:mod_sum} and Section~\ref{sec:model_selection}.

\section{Empirical Examples}
Every model fit reported in this section and in Section~\ref{sec:model_selection} is
preceded by \code{set.seed(1000)}, and unless stated otherwise uses the default MCMC
settings of \code{dsp\_fit()}. The replication script accompanying this article
reproduces each figure and printed value exactly.

\subsection{Smoothing Univariate Gaussian Time Series}\label{sec:g_smoothing_ex}
\begin{figure}[ht]
    \centering
    \begin{subfigure}[b]{0.45\textwidth}
      \includegraphics[width=\linewidth]{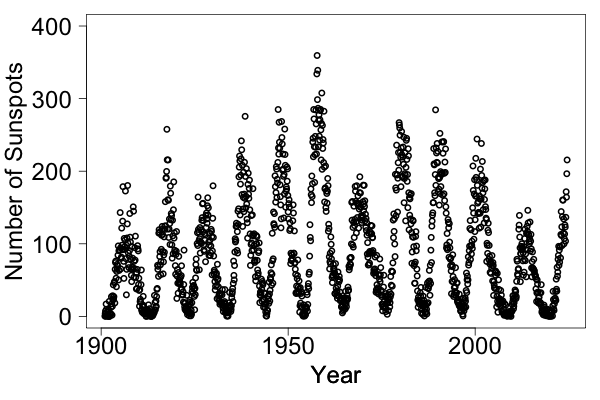}
      \caption{Observed Series $y_{t}$}
      \label{fig:btfasv_a}
    \end{subfigure}
        \begin{subfigure}[b]{0.45\textwidth}
      \includegraphics[width=\linewidth]{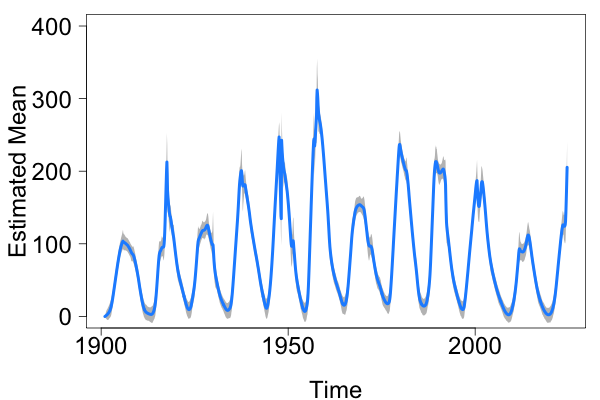}
      \caption{Estimated Mean}
      \label{fig:btfasv_b}
    \end{subfigure}
\caption{The monthly mean total sunspot numbers, defined as the average of daily total sunspot counts over each calendar month from January 1901 to August 2024, are shown in Figure~\ref{fig:btfasv_a}. A Bayesian Trend Filter with the dynamic horseshoe prior (BTF-DHS) is used to estimate the mean (Figure~\ref{fig:btfasv_b}). 95\% credible intervals are overlaid in grey.}
\end{figure}
Bayesian Trend Filter (BTF) from Section~\ref{sec:g_smoothing} can be applied to estimate the time-varying mean of a univariate Gaussian time series. We consider the monthly mean total sunspot numbers. The series contains the average daily sunspot count for each calendar month from January 1901 to August 2024, obtained from \citet{sunspot}.  As shown in Figure~\ref{fig:btfasv_a}, sunspot activity displays a well-known cyclic pattern associated with the 11-year solar cycle, which reflects periodic fluctuations in solar magnetic activity. The model is fitted using the following code:
\begin{Code}
R> spec <- dsp::dsp_spec(family = "gaussian",
                        model = "smoothing",
                        obsSV = "const",
                        D = 2,
                        evol_error = "DHS")
R> sunspot_fit <- dsp::dsp_fit(y,spec)
\end{Code}
Figure~\ref{fig:btfasv_b} is then produced by:
\begin{Code}
R> plot(sunspot_fit,"mu",alpha = 0.05,xlab = "Time", ylab = "Estimated Mean")
\end{Code}
The estimated mean process, $\{\mu_t\}_{t=1}^{T}$, shown in Figure~\ref{fig:btfasv_b}, captures the quasi-periodic structure well. We set $D = 2$ to encourage piecewise linear trend. In general, penalizing the $D$th-order difference promotes local polynomial structure of degree $D-1$. $D=1$ yields piecewise constant behavior, $D=2$ favors piecewise linearity, and higher-order differences induce increasingly smooth trends.

\subsection{Adaptive Stochastic Volatility}\label{sec:asv_ex}
\begin{figure}[ht]
    \centering
    \includegraphics[width=0.6\linewidth]{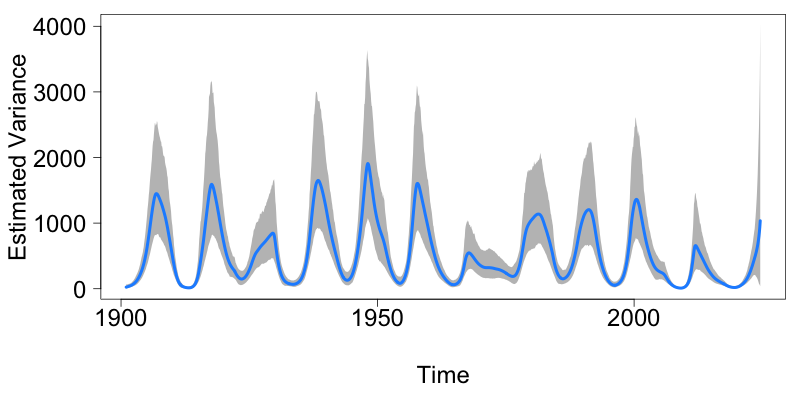}
   \caption{Estimated variance of the monthly mean total sunspot numbers from January 1901 to August 2024, obtained by fitting an Adaptive Stochastic Volatility (ASV) model (\code{obsSV = "ASV"}). The posterior mean and 95\% credible intervals are overlaid in grey.}\label{fig:btfasv_c}
\end{figure}
In addition to changes in the mean function, the observation variance may also evolve over time, which is the adaptive stochastic volatility model of Section~\ref{sec:asv}. The \pkg{dsp} package allows the mean and variance processes to be modeled jointly by combining the Bayesian trend filter with ASV. Consider the sunspot data example shown in Section~\ref{sec:g_smoothing_ex}. In addition to the cyclical behavior in the mean, Figure~\ref{fig:btfasv_a} displays changes in volatility. Periods of high sunspot activity tend to coincide with elevated volatility, while quieter phases exhibit substantially lower variance. To accommodate this heteroskedastic structure, the observation error variance can be modeled with ASV by replacing \texttt{obsSV = "const"} with \texttt{"ASV"} in the model specification:
\begin{Code}
R> spec_ASV <- dsp::dsp_spec(family = "gaussian",
                             model = "smoothing",
                             obsSV = "ASV",
                             D = 2,
                             evol_error = "DHS",
                             D_asv = 2)
R> sunspot_fit_asv <- dsp::dsp_fit(y,spec_ASV)
\end{Code}
The resulting posterior estimate of the variance process is shown in Figure~\ref{fig:btfasv_c} and is produced using the following command:
\begin{Code}
R> plot(sunspot_fit_asv,"obs_sigma_t2",alpha = 0.05,
        xlab = "Time", 
        ylab = "Estimated Variance")
\end{Code}
More generally, this observation-variance specification is not limited to Gaussian smoothing. The same \code{obsSV = "ASV"} option can also be combined with the time-varying regression model in Section~\ref{sec:g_regression} and the changepoint model in Section~\ref{sec:g_cp}. This adaptive stochastic volatility fit is the one carried forward to the posterior predictive examples in Section~\ref{sec:pp}.

\subsection{Changepoint Model}\label{sec:abco_ex}
\begin{figure}[ht]
    \centering
    \begin{subfigure}[b]{0.32\textwidth}
      \includegraphics[width=\linewidth]{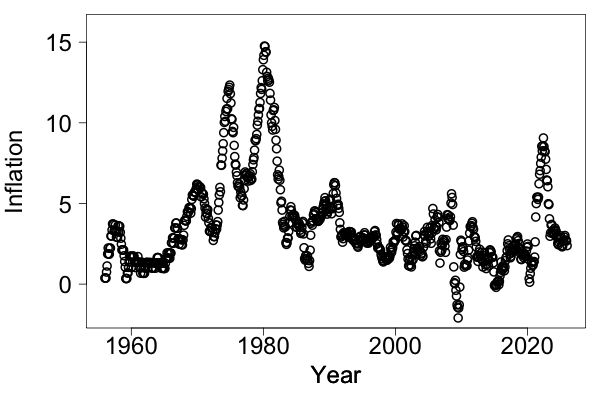}
      \caption{Observed Series $y_{t}$}
      \label{fig:abco_a}
    \end{subfigure}
        \begin{subfigure}[b]{0.32\textwidth}
      \includegraphics[width=\linewidth]{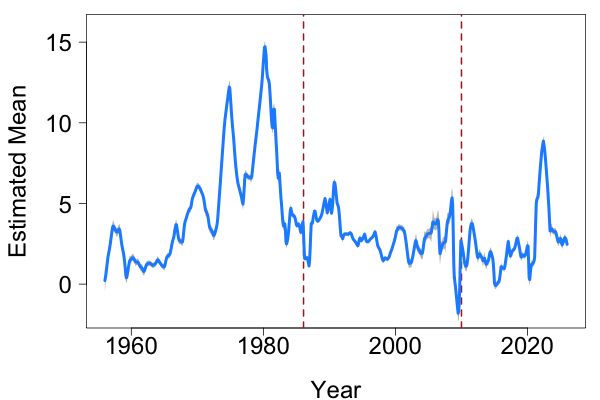}
      \caption{Mean with changepoints}
      \label{fig:abco_b}
    \end{subfigure}
    \begin{subfigure}[b]{0.32\textwidth}
      \includegraphics[width=\linewidth]{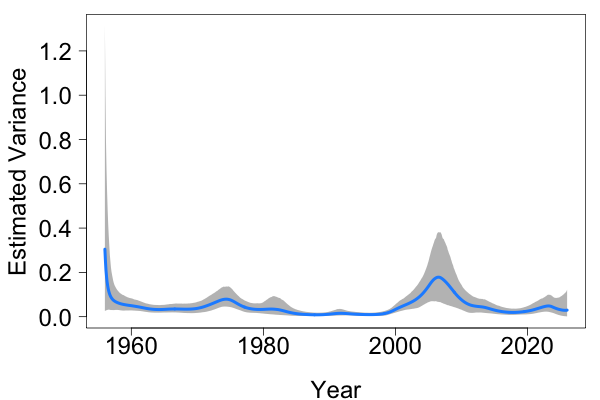}
      \caption{Variance}
      \label{fig:abco_c}
    \end{subfigure}
\caption{Monthly Consumer Price Index (CPI) from January 1956 to January 2026 is shown in Figure~\ref{fig:abco_a}. The Adaptive Bayesian Changepoints with Outliers (ABCO) is used to estimate the mean and changepoint locations as shown in Figure~\ref{fig:abco_b}, and the variance, $\sigma^2_{\epsilon,t}$, estimated using ASV along with its pointwise 95\% credible intervals, is shown in Figure~\ref{fig:abco_c}.}
\end{figure}
ABCO (Section~\ref{sec:g_cp}) is designed for changepoint detection in noisy time series where abrupt distributional shifts, isolated outliers, and heteroskedastic noise may be present. To illustrate its use, we apply ABCO with ASV, as described in Sections~\ref{sec:asv} and~\ref{sec:g_cp}, to the U.S. Consumer Price Index (CPI) from January 1956 to January 2026 \citep{cpi} to analyze monthly inflation, shown in Figure~\ref{fig:abco_a}. The model can be fit as:
\begin{Code}
R> spec <- dsp::dsp_spec(family = "gaussian",
                        model = "changepoint",
                        D = 2,
                        obsSV = "ASV", 
                        D_asv = 2)
R> model <- dsp::dsp_fit(y = y, model_spec = spec,
                        nsave = 5000, nburn = 5000)
\end{Code}

Figures~\ref{fig:abco_b} and~\ref{fig:abco_c} are produced using the following commands:
\begin{Code}
R> plot(model, "mu", alpha = 0.05, cp_thres = 0.8,
        xlab = "Year", ylab = "Estimated Mean")
R> plot(model, "obs_sigma_t2", alpha = 0.05,
        xlab = "Year", ylab = "Estimated Variance")
\end{Code}
The first command plots the estimated time-varying mean together with predicted changepoints, and the second plots the estimated variance from ASV. Changepoints are obtained by passing the fitted \code{dsp} object to \code{predict()} and specifying the threshold \texttt{cp\_thres} = 0.8; the default value is 0.5. In Figure~\ref{fig:abco_b}, the estimated changepoints are overlaid as red vertical lines. The estimated changepoints align with major macroeconomic episodes that plausibly shifted the level of U.S. inflation.

The changepoint around 1986 is consistent with a major oil price shock that passed through energy prices into headline inflation \citep{oil}. The changepoint in early 2010 coincides with the aftermath of the Great Recession, when many interacting factors, including the rebound from the 2009 deflationary episode and shifting inflation expectations, reshaped U.S. inflation dynamics \citep{ball2011,coibion2015}.

The simultaneous credible bands of Section~\ref{sec:bands} give a joint assessment of the increments:
\begin{Code}
R> credible_bands(model, pars = "omega")
Simultaneous 95% credible bands for omega (838 time points)
Global Bayesian p-value (minimum SimBaS): 0.73
The band includes zero at every time point.

Use as.data.frame() for the values at each time point.
\end{Code}
The two summaries answer different questions. The changepoint rule of \code{predict()} considers each increment separately, comparing it with the threshold of the changepoint model, whereas the band assesses all increments jointly and can be considerably more conservative. Here, \code{predict()} assigns posterior probabilities above 0.8 to two changepoints, but the global Bayesian $p$-value is 0.73 and the band includes zero at every time point. At a changepoint, the posterior of the increment can mix draws near zero with draws of a large jump. Its posterior standard deviation is then large relative to its mean, so the band includes zero even when most draws exceed the threshold.

\subsection{B-Spline}
\begin{figure}[ht]
    \centering
    \begin{subfigure}[b]{0.32\textwidth}
      \includegraphics[width=\linewidth]{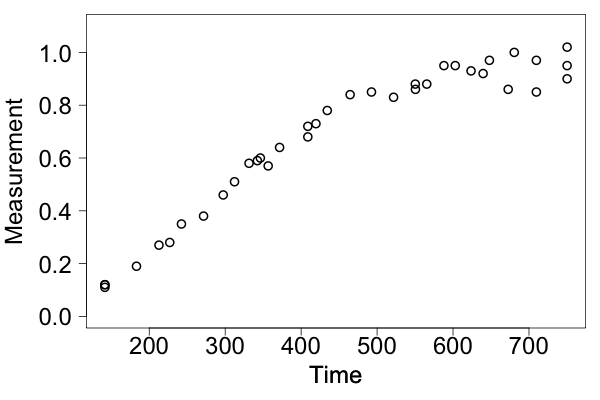}
      \caption{Observed Series $y_{t}$}
      \label{fig:bspline_a}
    \end{subfigure}
    \begin{subfigure}[b]{0.32\textwidth}
      \includegraphics[width=\linewidth]{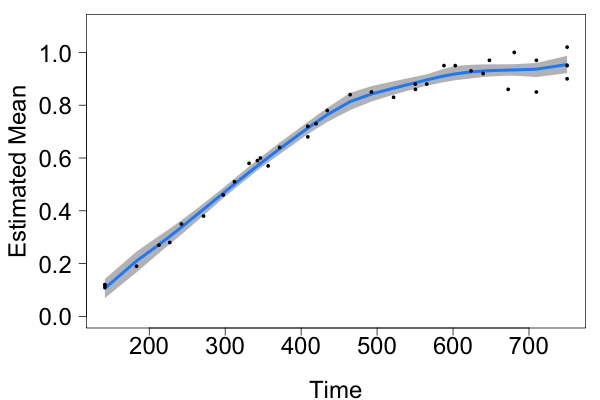}
      \caption{Mean (\texttt{num\_knots} = 20)}
      \label{fig:bspline_b}
    \end{subfigure}
    \begin{subfigure}[b]{0.32\textwidth}
      \includegraphics[width=\linewidth]{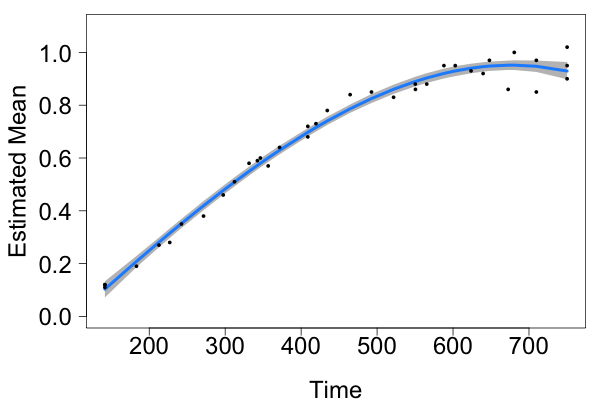}
      \caption{Mean (\texttt{num\_knots} = 4)}
      \label{fig:bspline_c}
    \end{subfigure}
\caption{Plant-organ measurements for group \texttt{RKW} from the \pkg{segmented} dataset~\citep{segmented} (Figure~\ref{fig:bspline_a}), and posterior trend estimates from the Gaussian B-spline DSP model with \texttt{num\_knots = 20} (Figure~\ref{fig:bspline_b}) and \texttt{num\_knots = 4} (Figure~\ref{fig:bspline_c}). The 95\% credible interval is drawn in grey.}
\end{figure}
Bayesian Trend Filter (BTF) from the previous section assumes evenly spaced observations. For irregularly spaced time series or functional data, the B-spline formulation in Section~\ref{sec:g_spline} can be used to estimate the time-varying mean. As an illustration, we use the plant-organ measurements for group \texttt{RKW} from the \pkg{segmented} package~\citep{segmented}, shown in Figure~\ref{fig:bspline_a}. In this implementation, the mean function is represented with a fourth-order (cubic) B-spline basis expansion.

The model can be fitted as:
\begin{Code}
R> spec <- dsp::dsp_spec(family = "gaussian",
                        model = "bspline",
                        times = times,
                        num_knots = 20)
R> model <- dsp::dsp_fit(y = y,model_spec = spec)
\end{Code}
There are two key arguments for the B-spline model: \texttt{times} and \texttt{num\_knots}.
The \texttt{times} argument provides the observation times. If it is not supplied, the model assumes equally spaced observations and uses a regular grid.

Figures~\ref{fig:bspline_b} and~\ref{fig:bspline_c} are produced using the following command:
\begin{Code}
R> plot(model,"mu",times = times,y_obs = y,
        alpha = 0.05,xlab = "Time", ylab = "Estimated Mean")
\end{Code}
The \texttt{num\_knots} argument controls the size of the B-spline basis (implemented with fourth-order, i.e., cubic, splines), so it must be at least 4. The default is \texttt{num\_knots = 20}. In practice, larger \texttt{num\_knots} gives a more flexible fit that can track local variation more closely, while smaller \texttt{num\_knots} gives a smoother global trend. The estimate with \texttt{num\_knots = 20} is shown in Figure~\ref{fig:bspline_b}, and the estimate with \texttt{num\_knots = 4} is shown in Figure~\ref{fig:bspline_c}; as expected, the former is more locally adaptive and the latter is smoother.

\subsection{Regression}\label{sec:count_reg}
\begin{figure}[ht]
    \centering
    \begin{subfigure}[b]{0.32\textwidth}
      \includegraphics[width=\linewidth]{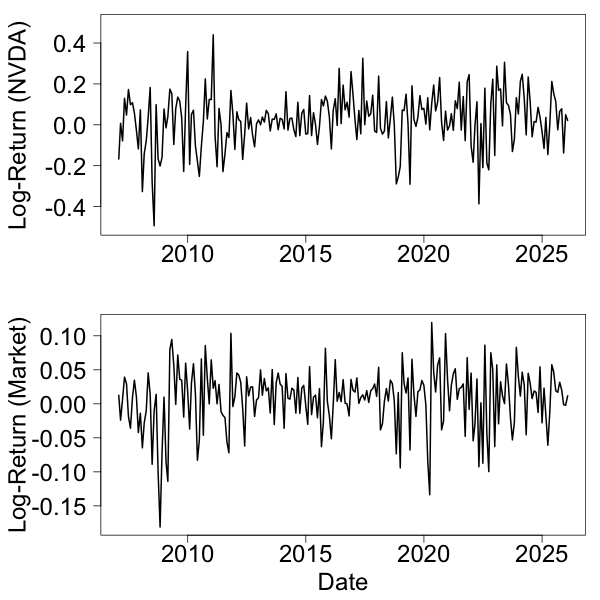}
      \caption{Log-Returns}
      \label{fig:reg_a}
    \end{subfigure}
    \begin{subfigure}[b]{0.32\textwidth}
      \includegraphics[width=\linewidth]{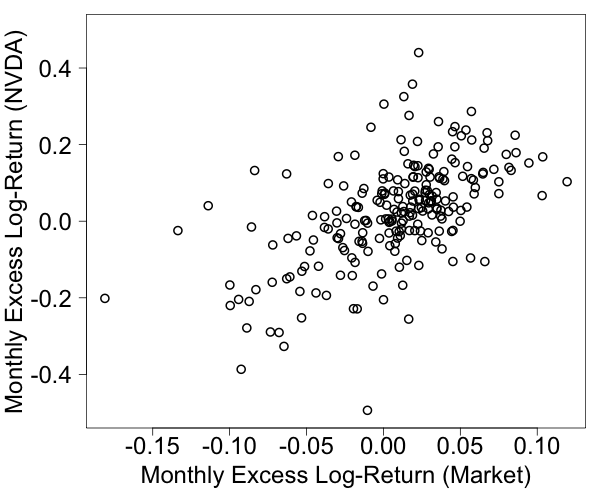}
      \caption{$R_{i,t} - R_{f,t}$ vs $R_{m,t} - R_{f,t}$}
      \label{fig:reg_b}
    \end{subfigure}
    \begin{subfigure}[b]{0.32\textwidth}
      \includegraphics[width=\linewidth]{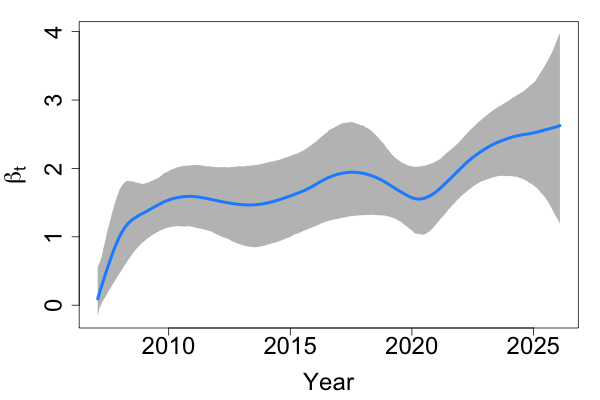}
      \caption{CAPM parameter}
      \label{fig:reg_c}
    \end{subfigure}
\caption{Monthly excess log-returns for NVDA (top) and the market proxy SPY (bottom), from January 2007 to January 2026, are shown in Figure~\ref{fig:reg_a}. Figure~\ref{fig:reg_b} shows the scatter plot of NVDA versus market excess returns, and Figure~\ref{fig:reg_c} reports the estimated time-varying Capital Asset Pricing Model (CAPM) loading with pointwise 80\% credible intervals drawn in grey.}
\end{figure}

Time-varying parameter regression, introduced in Section~\ref{sec:g_regression}, is useful when the strength of association between a response and its predictors is expected to evolve over time. As an illustration, we analyze monthly returns of NVIDIA (NVDA) stock in relation to the overall U.S. market (Figure~\ref{fig:reg_a}) through a time-varying Capital Asset Pricing Model (CAPM). Let $R_{i,t}$ denote the NVDA return at time $t$, $R_{f,t}$ the risk-free rate, and $R_{m,t}$ the market return. Under the zero alpha CAPM:
\begin{align*}
    R_{i,t}-R_{f,t}=\beta_t\,(R_{m,t}-R_{f,t})+\epsilon_t,
\end{align*}
where the market exposure ($\beta_t$) is allowed to change across market regimes.

Figure~\ref{fig:reg_b} shows a scatter plot of NVDA's monthly excess log-return ($R_{i,t}-R_{f,t}$) against the market excess log-return ($R_{m,t}-R_{f,t}$), indicating a clear positive association between NVDA and overall market movements. We fit the time-varying CAPM using:
\begin{Code}
R>  spec <-dsp::dsp_spec(family = "gaussian",
                         model = "regression",
                         X = X)
R>  model <-dsp::dsp_fit(y = y, model_spec = spec,
                         nsave = 5000, nburn = 20000)
\end{Code}
where $X$ is a $T \times p$ design matrix and $y$ is the response vector of NVDA excess log-returns. In this example, $p=1$, with the market excess log-return as the only covariate. For regression models, the predictor matrix $X$ must be supplied through \code{dsp\_spec()}.

The estimated time-varying loading $\hat{\beta}_t$ and its pointwise 80\% credible intervals are shown in Figure~\ref{fig:reg_c}, produced with the following command:
\begin{Code}
R> plot(model, "beta", alpha = 0.2, xlab = "Year", ylab = expression(beta[t]))
\end{Code}
We show 80\% intervals, following the convention in finance and macroeconomics of reporting intervals narrower than 95\% for time-varying coefficients, such as the 68\% intervals of \citet{primiceri2005}. The estimates suggest that $\hat{\beta}_t$ is roughly 1.5 prior to 2020, dips briefly around 2020, and then increases to around 2.0 by 2022, consistent with NVDA becoming more sensitive to market swings in the post-2020 period.

Since zero is a natural reference value for $\beta_t$, the simultaneous credible bands and SimBaS of Section~\ref{sec:bands} quantify the evidence for market exposure jointly over time:
\begin{Code}
R> credible_bands(model, pars = "beta", times = dates)
Simultaneous 95% credible bands for beta (229 time points)

Predictor 1:
Global Bayesian p-value (minimum SimBaS): 0.0054
The band excludes zero at 98 time points, in 3 periods:
      start        end length
 2008-12-31 2011-07-31     32
 2018-10-31 2019-12-31     15
 2020-10-31 2024-12-31     51

Use as.data.frame() for the values at each time point.
\end{Code}
where \code{dates} holds the month-end date of each observation. The global Bayesian $p$-value of 0.0054 indicates that the hypothesis $\beta_t = 0$ at every month is not supported by the data. The 95\% simultaneous band excludes zero in 98 of the 229 months, in the three periods listed. In the remaining months the band includes zero. Because the band must contain the entire trajectory in 95\% of the posterior draws, it is wider than pointwise intervals at the same credibility level, and wider still than the 80\% intervals in Figure~\ref{fig:reg_c}. Its inclusion of zero in these months does not by itself indicate that the exposure is zero.

\subsection{Smoothing Count Data}\label{sec:count_ex}
\begin{figure}[ht]
    \centering
    \begin{subfigure}[b]{0.32\textwidth}
      \includegraphics[width=\linewidth]{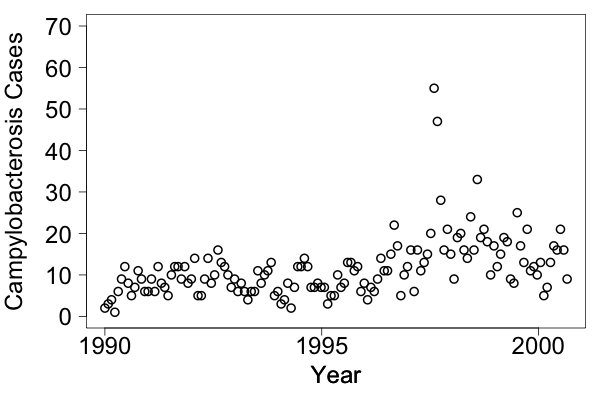}
      \caption{Observed Counts}
      \label{fig:count_a}
    \end{subfigure}
    \begin{subfigure}[b]{0.32\textwidth}
      \includegraphics[width=\linewidth]{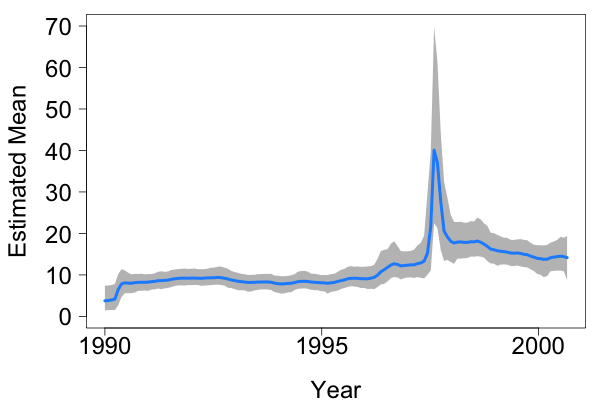}
      \caption{Mean (\texttt{D} = 1)}
      \label{fig:count_b}
    \end{subfigure}
    \begin{subfigure}[b]{0.32\textwidth}
      \includegraphics[width=\linewidth]{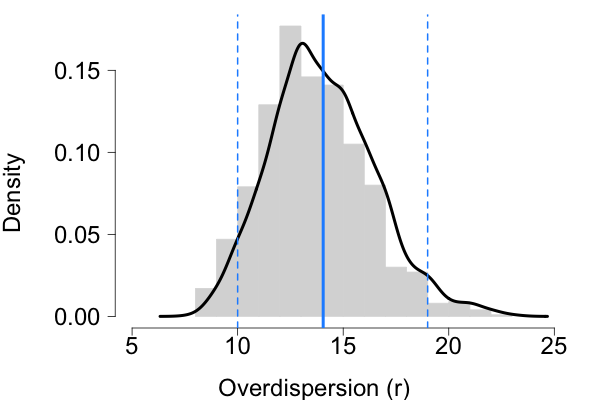}
      \caption{Overdispersion ($r$)}
      \label{fig:count_c}
    \end{subfigure}
\caption{Figure~\ref{fig:count_a} shows the number of campylobacterosis cases in northern Quebec, Canada. Figures~\ref{fig:count_b} and~\ref{fig:count_c} present the estimated conditional mean from the Negative Binomial Bayesian Trend Filter (NB-BTF) using first-order differencing (\texttt{D = 1}) and the overdispersion parameter $r$, respectively. The data are from \citet{countdata}. 95\% credible intervals are drawn in grey.}
\end{figure}
The Negative Binomial Bayesian Trend Filter (NB-BTF) from Section~\ref{sec:nb_smoothing} is designed for smoothing count-valued time series that exhibit overdispersion relative to the Poisson model. To illustrate its use, we analyze the \texttt{campy} data, which records the number of campylobacterosis cases in northern Quebec, Canada, observed every four weeks from January 1990 to the end of October 2000. The series contains 140 observations and is available in the \texttt{tscount} package \citep{countdata}. The model can be fit as:
\begin{Code}
R>  spec <- dsp::dsp_spec(family = "negbinomial",
                         model = "smoothing",
                         D = 1)
R>  model <- dsp::dsp_fit(y = y, model_spec = spec)
\end{Code}
The estimated conditional mean and overdispersion parameter $r$ are shown in Figure~\ref{fig:count_b} and~\ref{fig:count_c}, respectively, and are produced using the following commands:
\begin{Code}
R> plot(model,"mu",alpha = 0.05,xlab = "Year", ylab = "Estimated Mean")
R> plot(model,"r",alpha = 0.05,xlab = "Overdispersion (r)",ylab = "Density")
\end{Code}

The observed counts in Figure~\ref{fig:count_a} show substantial short-term variability, making the underlying mean difficult to assess directly. The estimated mean is relatively stable over much of the early and middle part of the series, roughly from 1990 to 1998, but displays a noticeable burst of change around 1997. Figure~\ref{fig:count_c} shows the posterior distribution of the overdispersion parameter \(r\). The concentration of posterior mass on moderate finite values of \(r\) indicates substantial overdispersion in the data.

\subsection{Posterior Predictive Draws}\label{sec:pp}
\begin{figure}[ht]
    \centering
    \begin{subfigure}[b]{0.45\textwidth}
      \includegraphics[width=\linewidth]{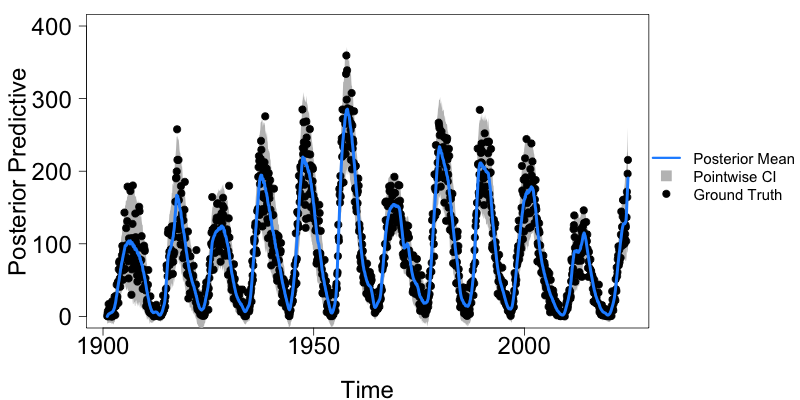}
      \caption{Without Missing Values}
      \label{fig:pred}
    \end{subfigure}
    \begin{subfigure}[b]{0.45\textwidth}
      \includegraphics[width=\linewidth]{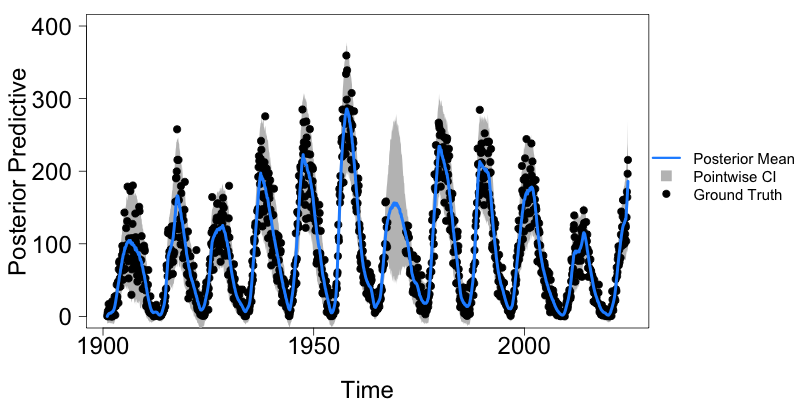}
      \caption{With Missing Values}
      \label{fig:pred_m}
    \end{subfigure}
    \caption{The posterior predictive mean and 95\% pointwise posterior predictive intervals, computed from the \texttt{ypred} draws, overlaid with the observed sunspot counts.  Panel (a) shows the model fit to the complete series. In panel (b), the observations from August 1967 through October 1971 are removed before fitting. The fitted model is the adaptive stochastic volatility specification from the example in Section~\ref{sec:asv_ex}.}
\end{figure}

The fitting function \texttt{dsp\_fit} returns draws from the posterior predictive distribution for all provided observation times, for every model in Section~\ref{sec:methods}. For each iteration of the MCMC, one random observation value per time point is sampled using the current model parameter values, which returns a full posterior predictive distribution at each time point. Therefore, the full posterior predictive distribution is available to construct point estimates, prediction intervals, or other quantities of interest (Figure~\ref{fig:pred}). The draws are stored as the parameter \texttt{ypred}, whose pointwise summaries are returned by \code{fitted(object, pars = "ypred")}. One use of the posterior predictive distribution is to impute missing values which can be achieved by providing missing values in the input vector \texttt{y}. As an artificial example, the following example replaces a block of observations from the sunspot series with \texttt{NA} and refits:
\begin{Code}
R> spec <- dsp::dsp_spec(family = "gaussian",
                         model = "smoothing",
                         obsSV = "ASV",
                         D = 2,
                         evol_error = "DHS",
                         D_asv = 2)
R> y[800:850] <- NA
R> sunspot_fit_missing = dsp::dsp_fit(y,spec)
R> plot(sunspot_fit_missing,"ypred",alpha = 0.05,
        true_values = y,
        xlab = "Time", 
        ylab = "Posterior Predictive")
\end{Code}
Here the observations from August 1967 through October 1971 are set to missing before fitting. The model is fit to the remaining data, and the posterior predictive draws at the missing time points yield an imputed mean and pointwise intervals over the gap, shown in Figure~\ref{fig:pred_m}. The prediction intervals over the gap are wider than those in the neighboring observed regions, reflecting the greater uncertainty where no concurrent observation informs the fit.
\begin{figure}[ht]
    \centering
    \begin{subfigure}[b]{0.45\textwidth}
      \includegraphics[width=\linewidth]{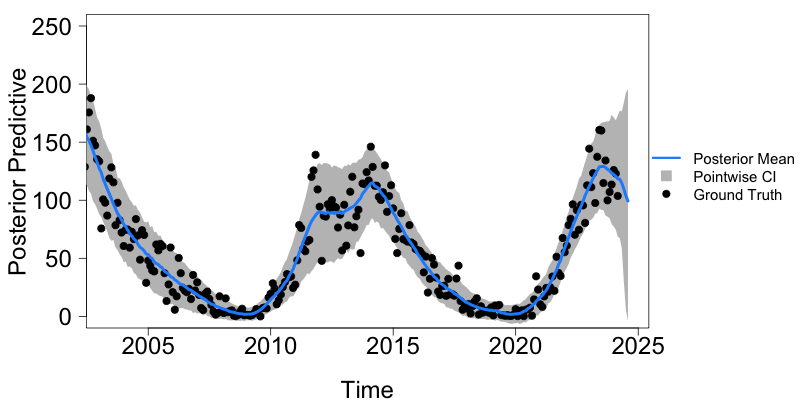}
      \caption{2003--2024}
      \label{fig:fut_full}
    \end{subfigure}
    \begin{subfigure}[b]{0.45\textwidth}
      \includegraphics[width=\linewidth]{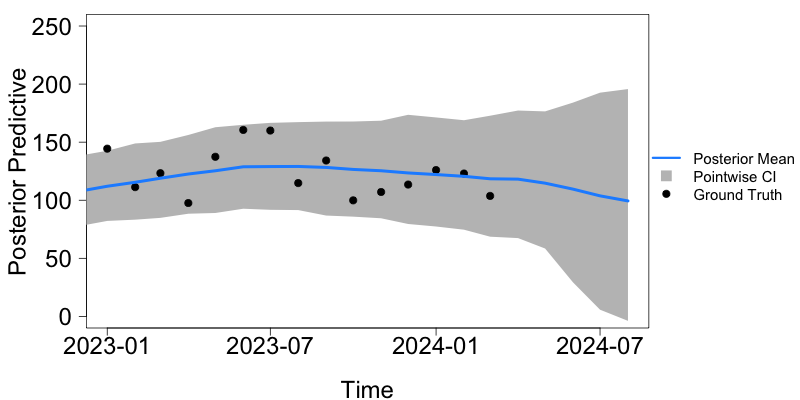}
      \caption{Forecast region}
      \label{fig:fut_zoom}
    \end{subfigure}
    \caption{Posterior predictive forecast for the sunspot series, using the adaptive stochastic volatility specification of Section~\ref{sec:asv_ex}, where the final five months are forecast. Panel (a) shows the series from May 2003 to August 2024, and panel (b) zooms on the forecast region. The forecast mean and 90\% pointwise prediction intervals are shown.}
    \label{fig:plot_fut}
\end{figure}

The same approach extends to forecasting: padding the observed series with missing values at the end yields posterior predictive draws at future time points. Figure~\ref{fig:fut_full} demonstrates this on the sunspot series, where the final five observations are set to missing before fitting. The model is fit to the remaining data, and the posterior predictive draws at these time points yield a forecast mean and pointwise prediction intervals. Figure~\ref{fig:fut_zoom} zooms on the forecast region, where the prediction intervals widen with the forecast horizon, reflecting the increasing uncertainty.

\subsection{Model Summaries}\label{sec:mod_sum}
The object examined in this subsection and in Section~\ref{sec:cpa} is the constant-variance fit of Section~\ref{sec:g_smoothing_ex}. The parameters with stored posterior draws are listed by \code{par_names()}, and the elements present depend on the chosen model and options, as described in Section~\ref{sec:output_des}. The \code{summary()} method then gives a compact overview of the fit:
\begin{Code}
R> par_names(object)
[1] "mu"            "ypred"         "evol_sigma_t2" "obs_sigma_t2" 
[5] "dhs_phi"       "dhs_mean"      "loglike"
R> summary(object)
Gaussian likelihood with Bayesian trend smoothing, constant error variance, and 
2 degrees of differencing.
Prior for the mean function: Dynamic horseshoe prior
Prior for the variance function: Normal-inverse-gamma prior

MCMC: 1000 saved draws (burn-in 1000, thinning interval 5); T = 1484 

Time-varying and vector parameters:
                      length posterior mean range
mu                      1484     [-0.1666, 311.8]
ypred                   1484     [-0.6802, 312.8]
evol_sigma_t2[1:2]         2       [7.554, 9.354]
evol_sigma_t2[3:1484]   1482    [0.09829, 384000]

Scalar parameters:
               mean      sd   2.5%  97.5% ess
obs_sigma_t2  447.8   22.25  411.4  485.1 201
dhs_phi      0.6412 0.06078 0.5003 0.7325  13
dhs_mean      -3.41  0.3934 -4.212 -2.692  24
loglike       -6635   18.79  -6673  -6603  36

DIC: 13422.2 (p_D = 152.4); with p_V: 14529.6 (p_V = 706.1)
\end{Code}
Since the observation variance is constant in time under \code{obsSV = "const"}, \texttt{obs\_sigma\_t2} is stored as one draw per iteration and summarized with the scalar parameters. The first two evolution variances, \texttt{evol\_sigma\_t2[1:2]}, belong to the initial states and are reported separately. The two versions of the DIC are returned as a data frame by \code{DIC()}, which is convenient when comparing fits, as in Section~\ref{sec:model_selection}, and the pointwise posterior summaries of a time-varying parameter are returned by \code{fitted()}:
\begin{Code}
R> DIC(object)
  estimator      DIC    p_eff
1       p_D 13422.19 152.4244
2       p_V 14529.61 706.1343
R> head(fitted(object, pars = "mu"))
           mean      2.5%         25%         50%        75%    97.5%
[1,] -0.1665924 -2.297058 -0.35833984 -0.03776632 0.08798273 1.358579
[2,]  0.1407658 -1.300522 -0.07824423  0.02347757 0.34850121 1.867078
[3,]  0.4719279 -1.736743 -0.11343171  0.28380118 0.91582398 3.411552
[4,]  0.8051193 -2.709073 -0.12938975  0.61502259 1.60175941 4.944937
[5,]  1.1330152 -3.127889 -0.17450288  0.93387194 2.22365583 6.486151
[6,]  1.5041704 -3.618585 -0.08327531  1.25487590 2.93166782 7.807058
\end{Code}

The posterior draws can also be coerced to a \pkg{coda} \code{mcmc} object, as described in Section~\ref{sec:coda}, making the convergence-diagnostic and plotting functions in \pkg{coda} directly available. For example, the effective sample size for the estimated mean function \texttt{mu} is obtained as follows:

\begin{Code}
# coerce posterior draws to a coda::mcmc object
R> mu_draws <- as.mcmc(object, pars = "mu")
R> head(coda::effectiveSize(mu_draws),5)
    mu[1]     mu[2]     mu[3]     mu[4]     mu[5] 
 991.8862  868.3669 1000.0000  887.5586  876.8419 
\end{Code}

\subsection{Custom Posterior Analysis}\label{sec:cpa}
The \pkg{dsp} package provides \code{summary()}, \code{fitted()}, \code{plot()}, and \code{predict()} methods for commonly used posterior summaries, visualization, and prediction. Analyses beyond these built-in methods can be carried out directly on the coerced posterior draws. The \code{mcmc} object obtained in Section~\ref{sec:mod_sum} has one row per posterior draw and one column per time point, so a user-defined functional of the latent process is evaluated by applying it across rows.

Consider the total variation of the mean process, $\mu$, $\operatorname{TV}(\mu)= \sum_{t=2}^{T} |\mu_t-\mu_{t-1}|,$ which summarizes the overall amount of temporal variation. Its posterior distribution is obtained by evaluating this functional separately for each posterior draw:
\begin{Code}
R> dim(mu_draws)
[1] 1000 1484
R> total_variation <- apply(mu_draws, 1, \(mu) sum(abs(diff(mu))))
R> mean(total_variation) # Posterior Mean
[1] 5028.561
R> quantile(total_variation, c(0.025, 0.5, 0.975)) # Posterior quantiles
    2.5%      50%    97.5% 
4672.518 5012.752 5473.081 
\end{Code}
More generally, users can apply their own functionals to each posterior draw to obtain posterior inference for features of the latent trajectory that are not provided by the package.

\section{Guidance on Model Specification}\label{sec:model_selection}
\begin{table}[ht]
\centering
\begin{tabular}{@{}l p{11.5cm}@{}}
\toprule
\textbf{Argument} & \textbf{Guidance} \\
\midrule
\texttt{family} &
\texttt{"gaussian"}: continuous-valued data \\
& \texttt{"negbinomial"}: count data \\[4pt]
\texttt{model} &
\texttt{"smoothing"}: trend only, no covariates \\
& \texttt{"bspline"}: irregular or functional data \\
& \texttt{"regression"}: time-varying covariates \\
& \texttt{"changepoint"}: structural breaks / outliers \\[4pt]
\texttt{D} &
\texttt{1}: piecewise-constant mean \\
& \texttt{2} (default): piecewise-linear mean \\[4pt]
\texttt{evol\_error} &
\texttt{"DHS"} (default): autocorrelated shrinkage; adapts to smooth and abrupt changes alike \\
& \texttt{"HS"}: same heavy tails as DHS, temporal independence; suited to isolated deviations \\
& \texttt{"BL"}: Laplace shrinkage; lighter-tailed, less locally adaptive \\
& \texttt{"SV"}: smooth Gaussian AR(1) scale; gradual drift, no sparsity \\
& \texttt{"NIG"}: constant scale over time; ridge/P-spline-like \\[4pt]
\texttt{obsSV} &
\texttt{"const"}: homoskedastic noise \\
& \texttt{"SV"}: smoothly evolving variance \\
& \texttt{"ASV"}: variance with abrupt regime shifts \\
\bottomrule
\end{tabular}
\caption{Quick-reference guide for selecting the family, model, and key \code{dsp\_spec()} arguments.}
\label{tab:combined-guide}
\end{table}
Most arguments of \code{dsp\_spec()} are not tuning parameters but model variations suited to different data contexts, such as count versus continuous-valued data or smooth versus abrupt changes in the mean. Table~\ref{tab:combined-guide} summarizes heuristics for choosing among them. In most applications, these heuristics, combined with knowledge of the data and visual inspection of the fit, are sufficient, and we do not recommend fitting many specifications and selecting among them on an information criterion alone. In this section, we illustrate this guidance on a simulated example, showing how each choice affects the fit.

\begin{figure}[ht]
    \centering
    \begin{subfigure}[b]{0.45\textwidth}
      \includegraphics[width=\linewidth]{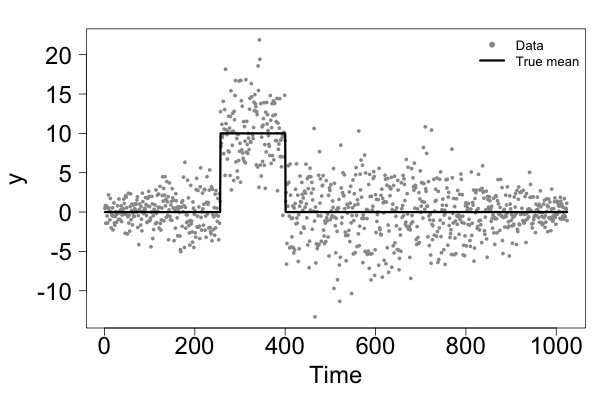}
      \caption{Observations (grey points) and true mean (black line)}
      \label{fig:simdata_mean}
    \end{subfigure}
    \begin{subfigure}[b]{0.45\textwidth}
      \includegraphics[width=\linewidth]{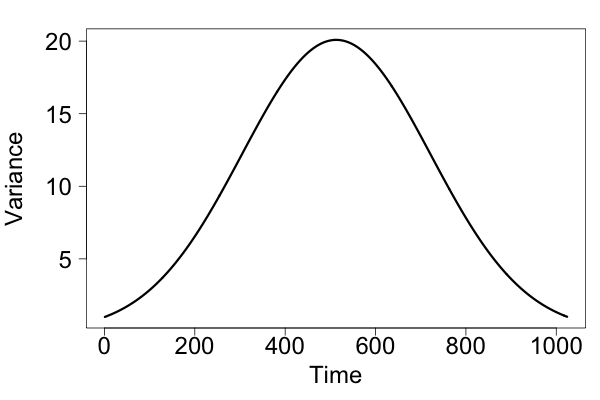}
      \caption{True variance process}
      \label{fig:simdata_var}
    \end{subfigure}
    \caption{Simulated data with a known mean and variance process: a piecewise-constant mean with two level shifts (Figure~\ref{fig:simdata_mean}) and a smoothly changing variance process (Figure~\ref{fig:simdata_var}), with $T = 1024$.}
    \label{fig:simdata}
\end{figure}

For this section, we consider simulated data generated using following commands:
\begin{Code}
R> set.seed(1000)
R> mu = dsp::simUnivariate(name = "levelshift")*10
R> sd = exp(dsp::simUnivariate(name = "quadratic")*1.5)
R> y = rnorm(1024,mean = mu,sd = sd)
\end{Code}
The series consists of $T = 1024$ observations generated as $y_t \sim N(\mu_t, \sigma_t^2)$, where $\mu_t$ is a piecewise-constant step function (Figure~\ref{fig:simdata_mean}) and $\sigma_t^2$ follows a smooth, symmetric process over time (Figure~\ref{fig:simdata_var}). Following Table~\ref{tab:combined-guide}, the continuous-valued response and our interest in the mean process suggest starting with the Gaussian smoothing model (\code{family = "gaussian"}, \code{model = "smoothing"}). To show how the remaining arguments affect the fit, we vary one component at a time while holding the others fixed, and compare the estimated mean or variance with the truth. Every fit in this section uses \code{nsave = 2000}, \code{nburn = 2000}, and \code{nskip = 4}, so that the comparisons are on equal footing. For reference, Table~\ref{tab:model-selection-dic} reports the DIC of each fit.

\begin{table}[ht]
\centering
\begin{tabular}{@{}llrr@{}}
\toprule
\textbf{Component varied} & \textbf{Specification} & \textbf{DIC} & $\boldsymbol{p_D}$ \\
\midrule
\code{obsSV}
    & \code{"const"} & 5280.3 & 26.5 \\
    & \code{"SV"}    & \textbf{5019.6} & 48.6 \\
    & \code{"ASV"}   & 5020.0 & 48.4 \\
\midrule
\code{D}
    & \code{1} & \textbf{4987.3} & 37.4 \\
    & \code{2} & 5019.6 & 48.6 \\
\midrule
\code{evol\_error}
    & \code{"DHS"} & \textbf{4987.3} & 37.4 \\
    & \code{"HS"}  & 4987.9 & 35.5 \\
    & \code{"BL"}  & 5050.0 & 88.1 \\
    & \code{"SV"}  & 4993.4 & 42.1 \\
    & \code{"NIG"} & 5061.2 & 94.1 \\
\bottomrule
\end{tabular}
\caption{Deviance information criterion (DIC) and effective number of
parameters $p_D$ for each fit in Section~\ref{sec:model_selection}. At each
stage, one component of the model specification is varied while the remaining
components are held fixed. Bold values mark the specification carried forward
to the subsequent stage.}
\label{tab:model-selection-dic}
\end{table}

\begin{figure}[ht]
    \centering
    \begin{subfigure}[b]{0.32\textwidth}
      \includegraphics[width=\linewidth]{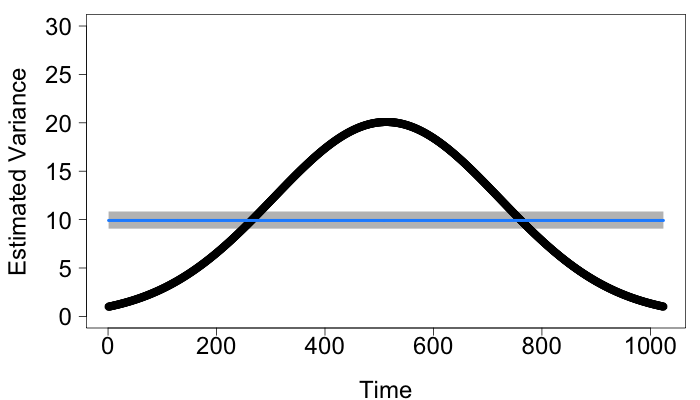}
      \caption{\code{obsSV = "const"}}
      \label{fig:obsSV_const}
    \end{subfigure}
    \hfill
    \begin{subfigure}[b]{0.32\textwidth}
      \includegraphics[width=\linewidth]{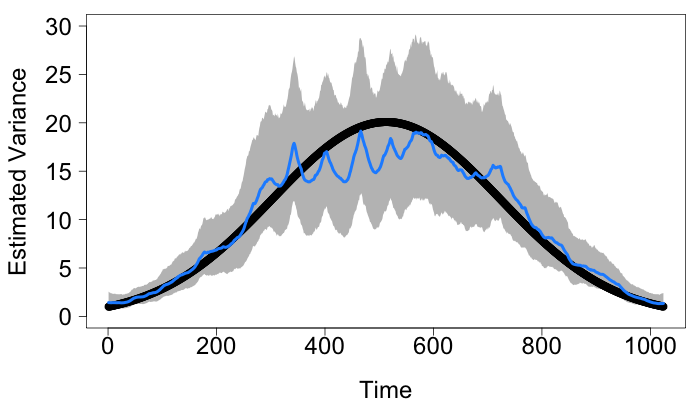}
      \caption{\code{obsSV = "SV"}}
      \label{fig:obsSV_sv}
    \end{subfigure}
    \hfill
    \begin{subfigure}[b]{0.32\textwidth}
      \includegraphics[width=\linewidth]{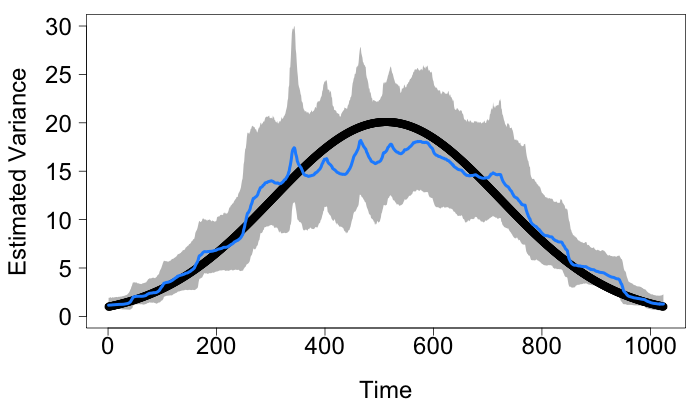}
      \caption{\code{obsSV = "ASV"}}
      \label{fig:obsSV_asv}
    \end{subfigure}
    \caption{Estimated observation variance under the three \code{obsSV}
    specifications: constant variance, stochastic volatility (SV), and
    adaptive stochastic volatility (ASV). The black line denotes the true
    variance process, the blue line denotes the posterior mean, and the
    shaded region denotes the 95\% credible intervals.}
    \label{fig:obsSV_comparison}
\end{figure}

We first consider the observation variance specification. As shown in Figure~\ref{fig:obsSV_comparison}, the constant model cannot capture the changing variance of the simulated series, whereas both \code{SV} and \code{ASV} recover its overall evolution. Because the true variance changes smoothly over time, Table~\ref{tab:combined-guide} suggests \code{SV}, which we carry forward. The DIC values in Table~\ref{tab:model-selection-dic} agree, with the constant-variance model clearly worse and \code{SV} and \code{ASV} nearly indistinguishable.

\begin{figure}[ht]
    \centering
    \begin{subfigure}[b]{0.45\textwidth}
      \includegraphics[width=\linewidth]{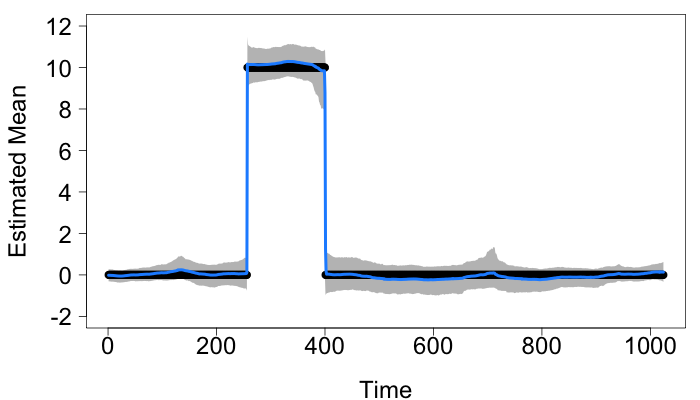}
      \caption{\code{D = 1}}
      \label{fig:D1}
    \end{subfigure}
    \hfill
    \begin{subfigure}[b]{0.45\textwidth}
      \includegraphics[width=\linewidth]{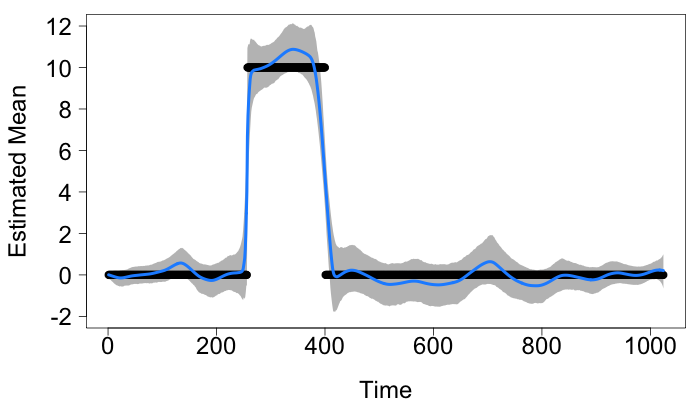}
      \caption{\code{D = 2}}
      \label{fig:D2}
    \end{subfigure}
    \caption{Estimated mean function under different degrees of differencing
    \code{D}. The black line denotes the true mean, the blue line denotes the
    posterior mean, and the shaded region denotes the 95\% credible intervals.}
    \label{fig:D_comparison}
\end{figure}

Holding \code{obsSV = "SV"} fixed, we next compare \code{D = 1} and \code{D = 2}. The choice of \code{D} determines the order of differencing applied to the latent mean process: \code{D = 1} favors piecewise-constant behavior, whereas \code{D = 2} favors piecewise-linear behavior. As shown in Figure~\ref{fig:D_comparison}, these choices lead to visibly different estimates of the mean. The first-order specification recovers the two level shifts sharply and remains flat elsewhere, whereas the second-order specification rounds the shifts and introduces spurious oscillations, consistent with the piecewise-constant structure of the true mean. The first-order specification also gives the lower DIC.

\begin{figure}[ht]
    \centering
    \begin{subfigure}[b]{0.32\textwidth}
      \includegraphics[width=\linewidth]{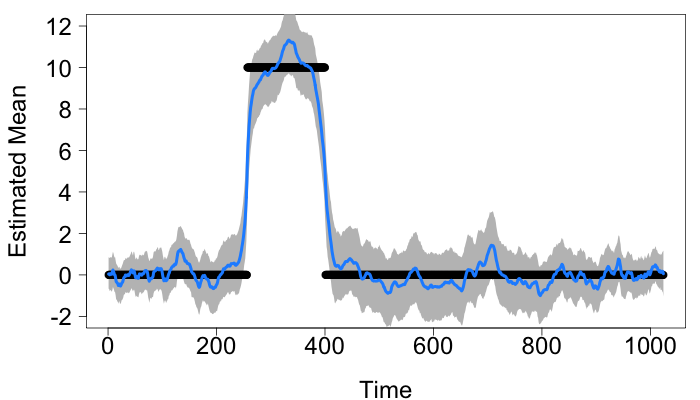}
      \caption{\code{evol\_error = "BL"}}
      \label{fig:prior_BL}
    \end{subfigure}
    \hfill
    \begin{subfigure}[b]{0.32\textwidth}
      \includegraphics[width=\linewidth]{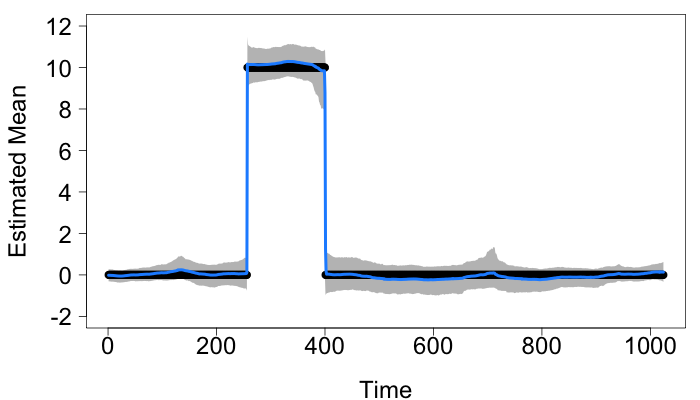}
      \caption{\code{evol\_error = "DHS"}}
      \label{fig:prior_DHS}
    \end{subfigure}
    \hfill
    \begin{subfigure}[b]{0.32\textwidth}
      \includegraphics[width=\linewidth]{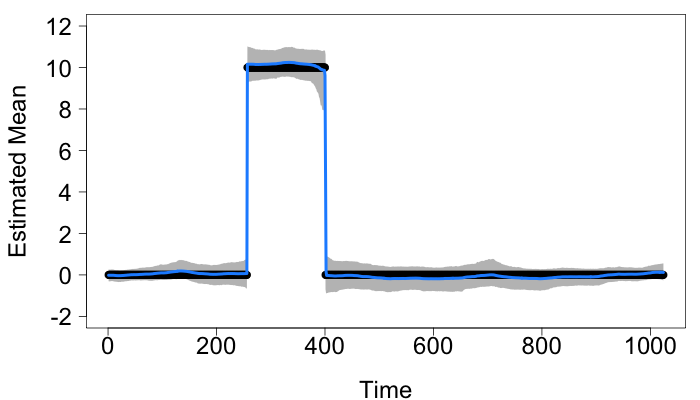}
      \caption{\code{evol\_error = "HS"}}
      \label{fig:prior_HS}
    \end{subfigure}

    \vspace{0.5em}

    \begin{subfigure}[b]{0.32\textwidth}
      \includegraphics[width=\linewidth]{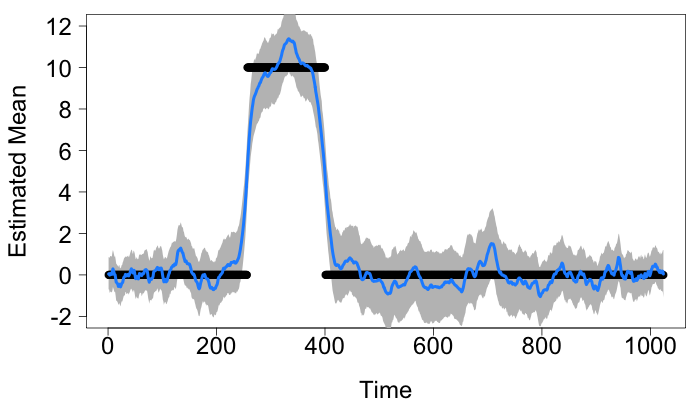}
      \caption{\code{evol\_error = "NIG"}}
      \label{fig:prior_NIG}
    \end{subfigure}
    \hspace{0.04\textwidth}
    \begin{subfigure}[b]{0.32\textwidth}
      \includegraphics[width=\linewidth]{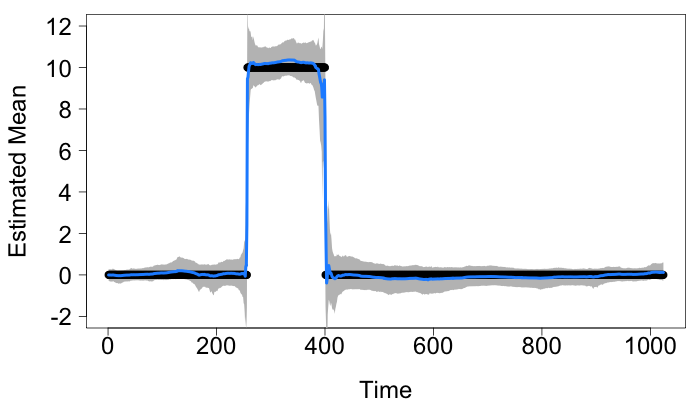}
      \caption{\code{evol\_error = "SV"}}
      \label{fig:prior_SV}
    \end{subfigure}
    \caption{Estimated mean function under different priors on the mean. The black line denotes the true mean, the blue line denotes the posterior mean, and the shaded region denotes the 95\% credible intervals.}
    \label{fig:prior_comparison}
\end{figure}
Finally, fixing \code{obsSV = "SV"} and \code{D = 1}, we compare the five available prior specifications for the mean. The shrinkage priors differ considerably in their ability to accommodate the long constant segments together with the two abrupt changes in the mean. As shown in Figure~\ref{fig:prior_comparison}, \code{DHS} and \code{HS} capture the level shifts without introducing spurious variation in the constant segments, whereas \code{BL} and \code{NIG} smooth over the shifts and produce noisier estimates of the constant segments. We therefore retain the default, \code{DHS}. The DIC values lead to the same conclusion, although the difference between \code{DHS} and \code{HS} is negligible.

More generally, DIC is a rough summary of the tradeoff between fit and complexity, and it does not replace visual inspection of the fit or convergence diagnostics. Lower values are preferred, but small differences, such as those between \code{SV} and \code{ASV} or between \code{DHS} and \code{HS} above, should not by themselves decide between specifications. In such cases, we suggest keeping the default or simpler specification unless the application or the fitted quantities indicate otherwise.

\pkg{dsp} also allows the posterior samples to be coerced for use with the \pkg{coda} package for further model diagnostics. For example, the effective sample size (ESS) can be computed as follows:
\begin{Code}
R> draws <- as.mcmc(final_fit, pars = "mu")
R> ESS = coda::effectiveSize(draws)
R> head(ESS,5)
   mu[1]    mu[2]    mu[3]    mu[4]    mu[5] 
1566.565 1535.486 1651.440 2000.000 1549.147 
\end{Code}
\begin{figure}[ht]
    \centering
    \begin{subfigure}[b]{0.45\textwidth}
      \includegraphics[width=\linewidth]{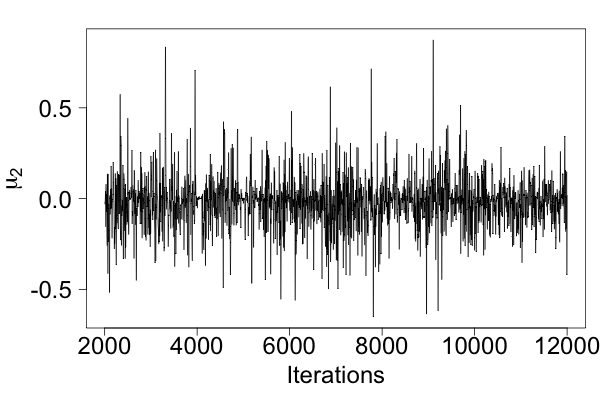}
      \caption{Trace plot of $\mu$ at $t =2$.}
      \label{fig:trace_mu2}
    \end{subfigure}
    \begin{subfigure}[b]{0.45\textwidth}
      \includegraphics[width=\linewidth]{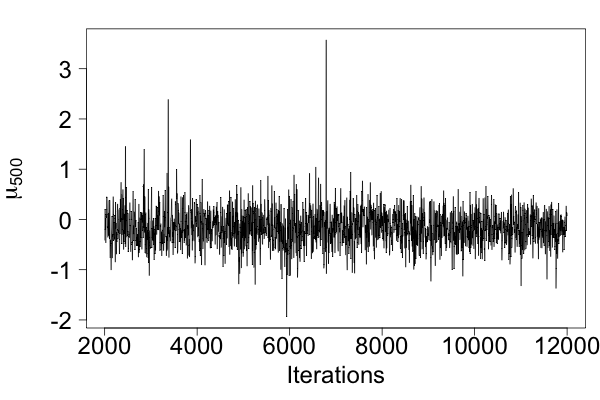}
      \caption{Trace plot of $\mu$ at $t = 500$.}
      \label{fig:trace_mu500}
    \end{subfigure}
    \caption{Trace plots of the estimated mean at two time points, $\mu_2$ (Figure~\ref{fig:trace_mu2}) and $\mu_{500}$ (Figure~\ref{fig:trace_mu500}).}
    \label{fig:traceplots}
\end{figure}
A trace plot shows the sampled values of a parameter across MCMC iterations, providing a visual check of mixing. Figure~\ref{fig:traceplots} show trace plots of the mean parameter at $t = 2$ and $t = 500$ and they are produced with the following commands:
\begin{Code}
R> coda::traceplot(draws[,2])
R> coda::traceplot(draws[,500])
\end{Code}
The horizontal axis reflects the underlying MCMC iteration index, including burn-in and thinning. \code{traceplot} and \code{effectiveSize} are two of several diagnostics that \pkg{coda} makes available through this coercion for further analysis.

In this section, the guidance in Table~\ref{tab:combined-guide}, together with visual inspection of the fits, led to a Gaussian smoothing model with \code{obsSV = "SV"}, \code{D = 1}, and \code{evol_error = "DHS"}. These choices agree closely with the known data-generating structure, which has a piecewise-constant mean with abrupt level shifts and a smoothly varying observation variance.

\section{Discussion}
This article introduced \pkg{dsp}, an \proglang{R} package for Bayesian time series modeling with DSP. The package provides a unified interface for fitting locally adaptive state-space models for Gaussian smoothing, count-data smoothing, time-varying regression, changepoint detection, and volatility estimation. By combining model specification, posterior computation, posterior predictive inference, summarization, and visualization within a common workflow, \pkg{dsp} makes DSP-based methods more accessible for users.

The package is also designed with computational practicality and extensibility in mind. The Gibbs sampling scheme is implemented in a modular way, which facilitates future development and makes it easier to incorporate new models, priors, and observation families. For a fixed model specification and number of MCMC iterations, the posterior sampling scales linearly with the length of the time series for the Gaussian models~\citep{dsp,abco,asv}. These features make \pkg{dsp} both a practical tool for Bayesian analysis of nonstationary time series and a foundation for further methodological and software development.

\section*{Computational details}

The results in this paper were obtained using \proglang{R}~4.5.3 with Apple \code{clang} version~21.0.0 for \proglang{C++}, the reference \pkg{BLAS} and \pkg{LAPACK} implementations distributed with \proglang{R}, running on macOS~26.7.1 with an Intel Core~i7-9750H processor. The packages most relevant to random number generation and to the linear algebra in the samplers were \pkg{RcppZiggurat}~0.1.8, \pkg{Rcpp}~1.1.1, \pkg{Matrix}~1.7-4, \pkg{spam}~2.11-3, \pkg{stochvol}~3.2.9, \pkg{BayesLogit}~2.3 and \pkg{coda}~0.19-4.1. \proglang{R} itself and all packages used are available from the Comprehensive \proglang{R} Archive Network (CRAN) at \url{https://CRAN.R-project.org/}.

The replication script provided as supplementary material reproduces every figure and printed value reported above. Each model fit is preceded by \code{set.seed(1000)}, and the sampler ties the separate random number stream used for the state draws to the stream that \code{set.seed()} controls, so repeated runs in separate \proglang{R} sessions on one machine return identical results. Complete session information is recorded at the end of the rendered script.

Results are not identical across platforms. The latent state is drawn through a sparse Cholesky factorization implemented with \pkg{Eigen} and compiled from source at installation. Different \proglang{C++} compilers make different, and individually valid, choices about instruction selection and the reassociation of floating-point expressions, so the factorization can differ in the final bits of the returned vector. Markov chain Monte Carlo amplifies such a difference, with the consequence that exact reproduction of the values reported here requires the platform described above. Under a different operating system or compiler, slightly different numbers are obtained. These are equivalent to our results according to the law of large numbers and lead to the same qualitative conclusions.

One consequence deserves explicit mention. The changepoints displayed in Figure~\ref{fig:abco_b} are those whose posterior probability exceeds \code{cp\_thres}, and the changepoint at February~1986 has posterior probability~0.80, which falls within Monte Carlo error of the threshold used. Whether it is displayed may therefore change across platforms. For this reason the replication script reports the posterior changepoint probabilities alongside the figure, and we recommend that users inspect those probabilities rather than relying on a thresholded summary alone.

\bibliography{ref.bib}
\end{document}